\documentclass{aa}

\usepackage{amssymb,times,graphicx, subfigure}
\usepackage[english]{babel}
\usepackage{txfonts}
\usepackage{xspace}
\usepackage{longtable}
\usepackage{lscape}
\usepackage{rotating}

\usepackage{natbib,afterpage,twoopt}
\usepackage[breaklinks=true]{hyperref}
\bibpunct{(}{)}{;}{a}{,}{,}

\def\vycma{VY~CMa\xspace}
\def\um{$\mu$m\xspace}

\def\msunyr{$M_{\sun}$\,yr$^{-1}$\xspace}
\def\kms{km\,s$^{-1}$\xspace}

\def\vlsr{\mbox{$\varv_{\mathrm{LSR}}$}\xspace}

\def\tioo{TiO$_2$\xspace}

\begin{document}

\title{ALMA observations of TiO$_2$ around VY Canis Majoris}
\titlerunning{\tioo around \vycma with ALMA}

\author{E. De Beck\inst{1}
        \and W. Vlemmings\inst{1}
        \and S. Muller\inst{1}
        \and J. H. Black\inst{1}
        \and E. O'Gorman\inst{1}
        \and A. M. S. Richards\inst{2}
        \and A. Baudry\inst{3,4}
        \and M. Maercker\inst{1}
        \and L. Decin\inst{5,6}
        \and E. M. Humphreys\inst{7}
}

\institute{
                        Department of Earth and Space Sciences, Chalmers University of Technology, Onsala Space Observatory, 43992 Onsala, Sweden         
                        \\ \email{elvire.debeck@chalmers.se}
        \and Jodrell Bank Centre for Astrophysics, School of Physics and Astronomy, University of Manchester, Manchester M13 9PL, UK
        \and Universit\'e de Bordeaux, LAB, UMR 5804, F-33270 Floirac, France
        \and CNRS, LAB, UMR 5804, F-33270 Floirac, France
        \and Instituut voor Sterrenkunde, Katholieke Universiteit Leuven, Celestijnenlaan 200D, 3001 Leuven, Belgium
        \and Sterrenkundig Instituut Anton Pannekoek, University of Amsterdam, Science Park 904, 1098 Amsterdam, The Netherlands 
        \and European Southern Observatory, Karl-Schwarzschild-Stra{\ss}e 2, 85748 Garching, Germany
}

\date{Received ------ ; accepted ------}

\abstract
{Titanium dioxide, \tioo, is a refractory species that could play a crucial role in the dust-condensation sequence around oxygen-rich evolved stars. To date, gas phase \tioo has been detected only in the complex environment of the red supergiant \vycma.}
{We aim to constrain the distribution and excitation of \tioo around \vycma in order to clarify its role in dust formation. }
{We analyse spectra and channel maps for \tioo extracted from ALMA science verification data.}
{We detect 15 transitions of \tioo, and spatially resolve the emission for the first time. The maps demonstrate a highly clumpy, anisotropic outflow in which the \tioo emission likely traces gas exposed to the stellar radiation field. An accelerating bipolar-like structure is found, oriented roughly east-west, of which the blue component runs into and breaks up around a solid continuum component. A distinct tail to the south-west is seen for some transitions, consistent with features seen in the optical and near-infrared.}
{We find that a significant fraction of \tioo remains in the gas phase outside the dust-formation zone and suggest that this species might play only a minor role in the dust-condensation process around extreme oxygen-rich evolved stars like \vycma.}

\maketitle


\defcitealias{ogorman2015_alma_vycma}{O+15}
\defcitealias{richards2014_alma_vycma}{R+14}
\defcitealias{kaminski2013_tio_tio2}{K+13a}
\defcitealias{kaminski2013_vycma_sma}{K+13b}

\section{Introduction}\label{sect:introduction}

The dust-formation sequence in the outflows of oxygen-rich evolved stars is not well understood. It is essential to address which gas-phase species provide the primary seeds. TiO$_2$ is considered an important seed refractory species with possibly higher nucleation rates than SiO \citep[e.g.][]{jeong2003_dustcondensation,lee2014_tio2_sio_nucleation}.  Moreover, presolar \tioo grains were tentatively identified by \cite{nittler1999_presolar_tio2grain}. Since SiO nucleation was recently indicated to be more relevant than previously thought under the relevant pressure and temperature conditions \citep{nuth2006_silicates,gail2013_silicates}, it is crucial to characterise the role of \tioo. The effect of non-stationarity on the nucleation is unknown. Shocks are, however, known to be present in the upper atmospheres of these evolved stars \citep[e.g.][]{chiavassa2011}.

Emission from gas phase \tioo has only been detected towards \vycma \citep[][hereafter \citetalias{kaminski2013_tio_tio2}, \citetalias{kaminski2013_vycma_sma}]{kaminski2013_tio_tio2,kaminski2013_vycma_sma}. 
The circumstellar environment of this red supergiant \citep[at 1.2\,kpc;][]{zhang2012_vycma_distance_vlba_vla} exhibits a high degree of morphological complexity, from optical to radio wavelengths and on spatial scales from a few to several thousand AU \citep[e.g.][]{humphreys2007_vycma_3dmorphology_kinematics,kaminski2013_vycma_sma,monnier2014_imagingbeautycontest,muller2007_vycma,shenoy2013_vycma_AO_2to5micron,smith2001_vycma,ziurys2007_vycma_complexity}. Recent ALMA observations spatially resolve H$_2$O maser emission, leading to the most accurate determination of the stellar position \citep[][hereafter \citetalias{richards2014_alma_vycma}]{richards2014_alma_vycma}. \citet[][hereafter \citetalias{ogorman2015_alma_vycma}]{ogorman2015_alma_vycma} describe the submillimeter continuum emission and report on a bright component south-east of the star, indicative of anisotropic mass loss.

\tioo is expected to be consumed by the nucleation process, but the detection of \tioo emission by  \citetalias{kaminski2013_tio_tio2} already suggested that a significant amount could possibly survive the dust formation. \citetalias{kaminski2013_tio_tio2} did not spatially resolve \tioo emission  at angular resolutions $\gtrsim$\,$1\arcsec$. The ALMA observations now allow us to  characterise the emission and the role of \tioo in the dust-formation process in more detail. 


\begin{table*}
        \centering
        \caption{Spectral coverage of the ALMA observations.}\label{tbl:spws}
        \begin{tabular}{ccccc}
                \hline\hline\\[-2ex]
                Setting & Frequency range & Resolution  &Rms noise      &Beam\\
                &(GHz)                          & (MHz)& (mJy\,beam$^{-1}$)&($\arcsec\times\arcsec,\mathrm{P.A.}$)\\\hline\\[-2ex]
                321&$309.501-310.437$&  0.98&   3&$0.23\times0.13,28.27^{\circ}$\\
                321&$310.501-311.437$&  0.98    &       2&$0.23\times0.13,28.48^{\circ}$\\
                325&$312.035-313.908$&  1.9     &       3&$0.24\times0.13,29.48^{\circ}$\\
                321&$320.725-321.662$&  0.98    &       3&$0.22\times0.13,27.31^{\circ}$\\
                321&$321.986-322.923$&  0.98    &       3&$0.22\times0.13,27.55^{\circ}$\\
                325&$324.181-326.054$&  1.9     &       6&$0.22\times0.12,29.27^{\circ}$\\
                658&$657.002-658.873$&  3.9     &       15&$0.11\times0.06,29.89^{\circ}$\\\hline
                \end{tabular}
                \tablefoot{Setting names refer to those used for the released ALMA CSV data, focussing on the H$_2$O masers at 321\,GHz, 325\,GHz, and 658\,GHz. }
\end{table*}    

\section{Observations}\label{sect:observations}
We retrieved ALMA science verification data on \vycma from the ALMA archives. The observations and data calibration and reduction are described by \citetalias{richards2014_alma_vycma}.  Table~\ref{tbl:spws} shows the spectral coverage and representative rms noise values for the six spectral windows in ALMA's band 7 ($\sim$0.9\,mm; $\sim$320\,GHz) and one in band 9 ($\sim$0.45\,mm; $\sim$660\,GHz). With projected baselines of 14\,m up to 2.7\,km, the spatial resolution at $\sim$320\,GHz and 658\,GHz is $\sim$\,$0\farcs2$ and  $\sim$\,$0\farcs1$, respectively, and the maximum recoverable scales are 8\farcs3 and 4\farcs0. The synthesised beam sizes are obtained using natural weighting.

The data reduction and quality of the continuum emission images at 321\,GHz and 658\,GHz are discussed in detail by \citetalias{richards2014_alma_vycma} and \citetalias{ogorman2015_alma_vycma}. No imaging artefacts are expected to arise from the array configuration owing to the excellent coverage of the visibility plane. 

\begin{figure*}
\centering
        \includegraphics[width=\linewidth]{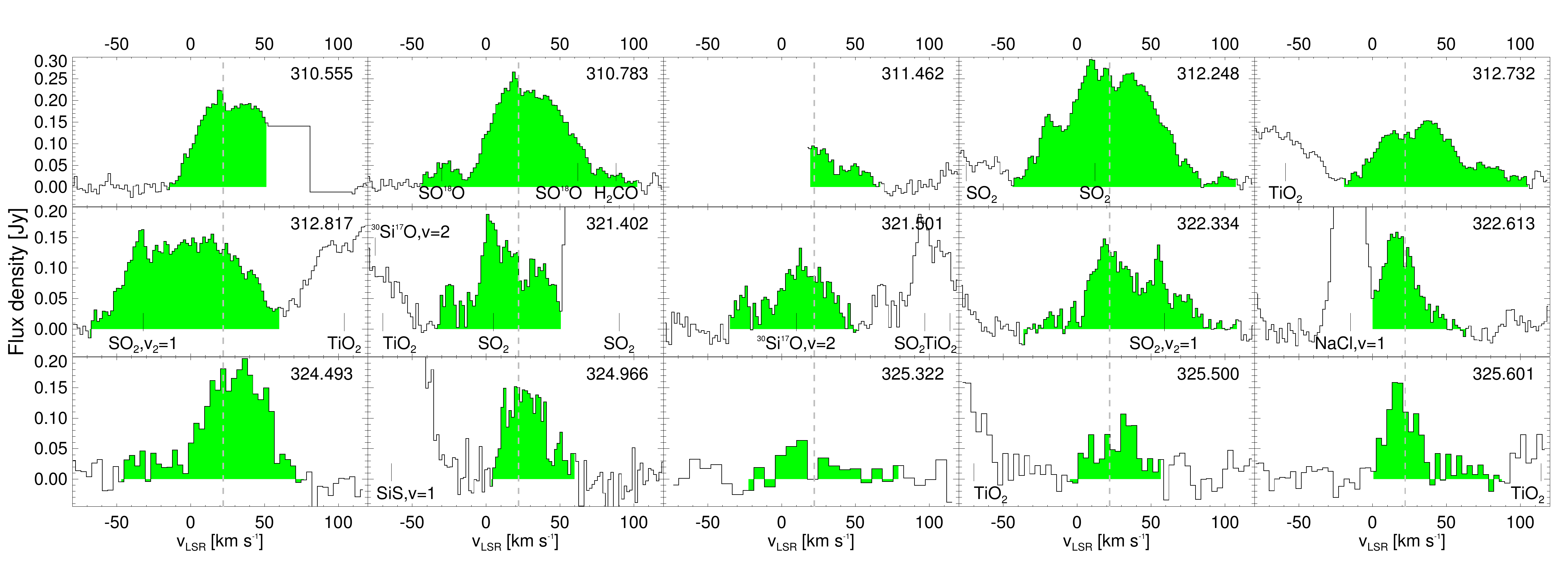}
        \caption{TiO$_2$ spectra extracted for a 1\arcsec\ diameter aperture around the stellar position. The vertical dashed lines indicate the stellar \vlsr of 22\,\kms, the shaded areas the \vlsr-ranges from Table~\ref{tbl:lineID}. We indicate identifications of species other than TiO$_2$ in the panels.}\label{fig:tio2all}
\end{figure*}

We note that the spectral window $312-314$\,GHz suffered from poor continuum subtraction owing to line crowding and a consequent lack of line-free spectral range. This causes a fraction of the continuum emission to leak into the channel maps of the \tioo lines at $\sim$312\,GHz (Figs.~\ref{fig:312248}, \ref{fig:312732}, \ref{fig:312817}). The emission seen at the position of clump C can be entirely attributed to the continuum itself, i.e. we can rule out molecular contribution at this position. The lines presented in Fig.~\ref{fig:tio2all} and Table~\ref{tbl:lineID}, and consequently also the reported peak and integrated intensities, were corrected for this effect on a line-to-line basis.

The imaged \tioo lines in the $312-314$\,GHz spectral window also show a contribution north-north-east and south-south-west of the star. These features are artefacts of the cleaning procedure likely caused by imperfect phase corrections on a number of intermediate baselines. The phase corrections were transferred from the self-calibrated 325\,GHz maser line located in the atmospheric absorption region.


\section{Results}\label{sect:results}
We analysed spectra extracted for a 1\arcsec\ diameter region around the stellar position. We show below that no \tioo emission is detected beyond this aperture. \tioo identifications are based on the Cologne Database for Molecular Spectroscopy \citep[CDMS;][]{mueller2001_cdms,mueller2005_cdms,bruenken2008_tio2}.  We detect 15 lines with upper-level energies $E_{\mathrm{up}}/k$ in the range $48-676$\,K and signal-to-noise ratios $S/N\approx5-17$ at velocity resolutions $0.9-7.6$\,\kms  (Table~\ref{tbl:lineID}, Fig.~\ref{fig:tio2all}). Of the \tioo lines detected by \citetalias{kaminski2013_tio_tio2,kaminski2013_vycma_sma} only those at 310.55\,GHz and 310.78\,GHz are observed with ALMA. The peak fluxes of the ALMA and SMA detections are consistent within the uncertainties. Moreover, of the other 13 lines of \tioo detected with ALMA, none were detected in the SMA survey, owing to a noise level in the ALMA data that was approximately 10 times lower. We detect no \tioo emission at $\sim$\,660\,GHz owing to the higher rms noise (Table~\ref{tbl:spws}). We detect no isotopic variants of Ti, consistent with the solar isotopic abundance ratios.
 
We expect no large flux losses in the ALMA observations of \tioo, given the $\sim$8\arcsec\/ recoverable scale at $\sim$320\,GHz. The presence of large-scale emission with low surface brightness cannot be entirely excluded, but is unlikely and will not change our main conclusions.

\begin{table*}[ht]
        \centering
        \caption{Overview of detected TiO$_2$ lines. Parameters listed for spectrum extracted in a $1\arcsec$ diameter aperture centred on the star. See also Fig.~\ref{fig:tio2all}. }\label{tbl:lineID}
\begin{tabular}{crrrrrrrrrc}
\hline\hline\\[-2ex]
Transition      & \multicolumn{1}{c}{$\nu_{\mathrm{lab}}$} & \multicolumn{1}{c}{$E_{\mathrm{up}}/k$} &  $\varv_{\mathrm{min}}$ &  $\varv_{\mathrm{max}}$ & \multicolumn{1}{c}{$\Delta\varv$} & \multicolumn{1}{c}{Rms} &  \multicolumn{1}{c}{Peak}   &\multicolumn{1}{c}{$I$}& Size    & SW \\
$J^{\prime}_{K^{\prime}_a,K^{\prime}_c} - J_{K_a,K_c}$& \multicolumn{1}{c}{(MHz)}   & \multicolumn{1}{c}{(K)}                 &\multicolumn{3}{c}{(\kms)}    & \multicolumn{1}{c}{(mJy)}& \multicolumn{1}{c}{(Jy)}         &\multicolumn{1}{c}{(Jy\,km\,s$^{-1}$)}& (\arcsec)& tail \\
                \hline\\[-2ex]
                $22(1,21)-21(2,20)$     &       310554.735      &       180.5                   &$-15.2$        &       51.3    &       1.0     &       16      &       0.22    &       9.57    &0.9&Y\\
                $23(1,23)-22(0,22)$     &       310782.713      &       182.4                   &$-43.9$        &       102.3   &       1.0     &       16      &       0.27    &       15.28   &0.9&Y\\
                $40(8,32)-39(9,31)$     &       311462.082      &       675.9                   &$19.4$ &       66.7    &       1.9     &       10      &       0.10    &       2.31    &0.8    &--\\
                $7(5,3)-6(4,2)$                 &       312248.341      &       47.9                    &$-43.7$        &       80.96   &       1.9     &       16      &       0.29    &       20.32   &0.8    &N\\
                $10(4,6)-9(3,7)$                &       312732.066      &       57.7                    &$-19.4$        &       105.1   &       1.9     &       16      &       0.15    &       9.43    &0.5&?\\
                $30(3,27)-30(2,28)$     &       312816.809      &       354.4                   &$-68.2$        &       60.0    &       1.9     &       16      &       0.16    &       12.98   &       0.5&?\\
                $11(4,8)-10(3,7)$               &       321401.936      &       65.7                    &$-33.9$        &       50.5    &       0.9     &       24      &       0.19    &       7.24    &       0.8     &?\\
                $28(2,26)-28(1,27)$     &       321501.043      &       298.8                   &$-35.0$        &       51.2    &       0.9     &       24      &       0.13    &       4.76    &0.9    &N\\
                $35(8,28)-35(7,29)$     &       322333.594      &       532.4                   &$-36.4$        &       108.1   &       1.0     &       19      &       0.15    &       7.03    &0.8    &N\\
                $33(8,26)-33(7,27)$     &       322612.696      &       481.4                   &$-0.1$ &       63.9    &       1.0     &       18      &       0.16    &       4.18    &0.9    &N\\
                $23(2,22)-22(1,21)$     &       324492.930      &       196.1                   &$-47.9$        &       79.4    &       1.9     &       29      &       0.21    &       9.64    &$\lesssim0.5$  &N\\
                $37(8,30)-37(7,31)$     &       324965.693      &       586.5                   &$3.6$  &       59.8    &       1.9     &       30      &       0.15    &       4.92    &0.5    &N\\
                $26(1,25)-26(0,26)$     &       325322.035      &       246.5                   &$-25.2$        &       81.8    &       7.6     &       21      &       0.11    &       1.85    &0.5    &N\\
                $26(4,22)-25(5,21)$     &       325500.415      &       281.5                   &$-7.7$ &       55.7    &       3.8     &       28      &       0.13    &       2.75    &0.5    &N\\
                $28(8,20)-28(7,21)$     &       325600.792      &       367.0                           &$-0.4$ &       90.2    &       3.8     &       28      &       0.18    &       3.89    &0.5    &N\\
                \hline\\[-2ex]
        \end{tabular}
        \tablefoot{Columns are transition, rest frequency, upper-level energy, minimum and maximum \vlsr reached, velocity resolution and rms noise at which the identification was made, peak and integrated flux, maximum diametric size of the integrated emission at $3\sigma$ level, and a marker for emission in the south-west tail (Y, N, ?, and -- for yes, no, maybe, and unknown). Species other than \tioo detected inside the plotted spectral windows are marked in Fig.~\ref{fig:tio2all}.}
\end{table*} 
 
Blending, proximity to strong lines, line crowding, and intrinsic line-shape irregularities complicate the  identification of \tioo emission. The lines at 310.55\,GHz and 311.46\,GHz are only partially covered in the observations; we believe the former to be a firm detection and the latter to be tentative. In both cases, no other candidate could be identified. Uncertainties on the relevant \tioo frequencies are $<2$\,MHz, except for the transition at 311.462\,GHz, where it is 5.555\,MHz \citep{bruenken2008_tio2}.

The \tioo lines detected with ALMA exhibit broad line profiles, as do those detected by \citetalias{kaminski2013_tio_tio2}. Assuming\footnote{See \citetalias{kaminski2013_vycma_sma} for a discussion of \vlsr.} a stellar \vlsr  of 22\,\kms, we find that the emission is very asymmetric in velocity space, ranging typically from $\sim$\,$-15$\,\kms to $\sim$\,$60$\,\kms  (Fig.~\ref{fig:tio2all}). The apparent extension of the blue wing of some lines to $\sim$\,$-45$\,\kms is likely coincidental and due to line blending, but we cannot exclude an actual \tioo contribution. Blend candidates are indicated in Fig.~\ref{fig:tio2all} and Table~\ref{tbl:lineID}. The lines with $E_{\mathrm{up}}/k=48$\,K\footnote{Continuum subtraction is hampered by the high line density. The red wing of the line is affected, but still shows the feature they have in common.}, 58\,K, 182\,K, 532\,K exhibit features in their red wings that extend to $\vlsr\approx105$\,\kms. A similar high-velocity outflow is also visible for some lines presented by \citetalias{kaminski2013_tio_tio2}.


\begin{figure*}
\includegraphics[width=\linewidth]{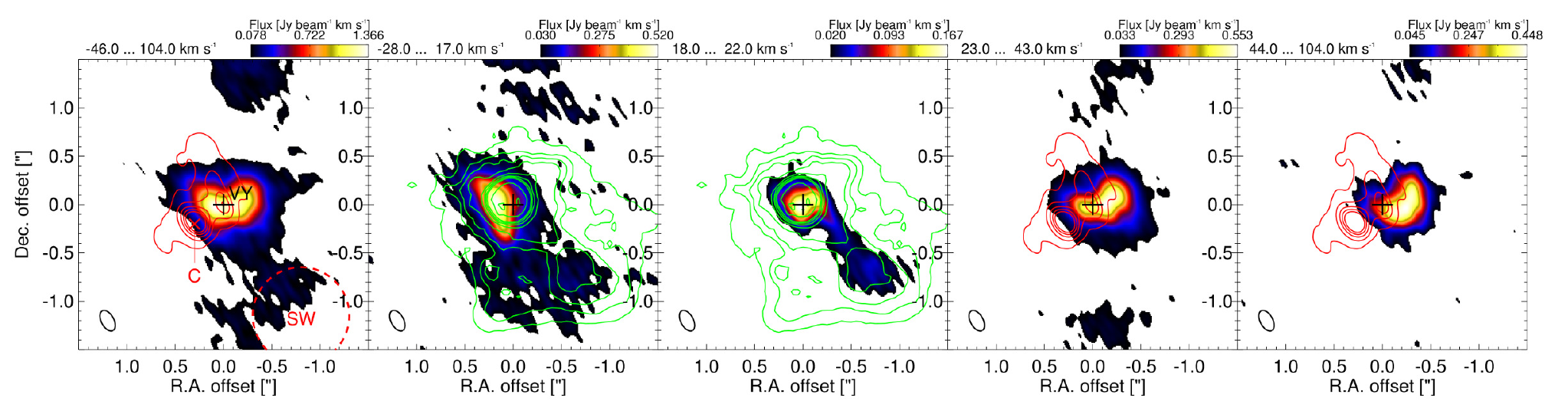}
\caption{TiO$_2$ morphology. Colour maps of emission at 310.78\,GHz integrated over the \vlsr-ranges indicated at the top left of each panel, cut off at $3\sigma$. Red contours show the 321\,GHz continuum at $[3,20,40,60,80]\sigma$; green contours show HST emission at $[3,5,7,10,20,30,40,50,100,200] \sigma$ \citep[][]{smith2001_vycma}. In the first panel we mark the position of the star (+, VY; black) and of the continuum component (x, C; red) to the south-east \citepalias{ogorman2015_alma_vycma,richards2014_alma_vycma}, and the position and approximate extent of the south-west clump \citep[SW, dashed 1\arcsec\ diameter circle; red][]{shenoy2013_vycma_AO_2to5micron}. The apparent north-south emission is thought to arise from dynamic-range limitations in the peak channels.} \label{fig:TiO2_fullstructure}
\end{figure*}

\subsection{Spatial distribution}\label{sect:morphology}
The detected \tioo emission is, for the most part, spatially resolved by the ALMA observations  with a $0\farcs23\times0\farcs13$ beam (see Figs.~\ref{fig:310554}--\ref{fig:mom0_all}), but several transitions show spatially unresolved emission at a 2\,\kms velocity resolution. We note a very complex behaviour of the emission peaks, with some only appearing  in one channel and not in the neighbouring channels. The often spatially unresolved ($\lesssim$145\,AU) peaks in the different velocity channels of all lines imply a clumpy and/or anisotropic wind.  The maps for 310.55\,GHz ($E_{\mathrm{up}}/k=181$\,K) and 310.78\,GHz ($E_{\mathrm{up}}/k=182$\,K) reveal that these two lines behave almost identically. Corresponding velocity channels (Figs.~\ref{fig:310554}, \ref{fig:310783}) trace roughly the same regions of the circumstellar environment and exhibit similar intensities, implying that these two lines are very tightly coupled in their excitation. Figure~\ref{fig:TiO2_fullstructure} shows the morphology of the TiO$_2$ emission at 310.78\,GHz and the positions of the star (VY), the continuum component C \citepalias{ogorman2015_alma_vycma,richards2014_alma_vycma}, and the south-west clump detected  at $\lambda\sim1-5$\,\um at $\sim$\,1\arcsec\ from the star \citep{smith2001_vycma,humphreys2007_vycma_3dmorphology_kinematics,shenoy2013_vycma_AO_2to5micron}. 

Given the similar energy levels, quantum numbers, and Einstein-$A$ coefficients of the transitions at 310.55\,GHz, 310.78\,GHz, and 324.49\,GHz, one expects similar line intensities and spatial distributions. However, although the channel-to-channel peaks of the 324.49\,GHz emission (Fig.~\ref{fig:324493}) correspond quite well to those of the other two, which strengthens its identification as \tioo, its intensity is clearly lower. It is hard to explain this discrepancy, but we note that as a consequence of the sensitivity of the atmospheric transmission at $\sim$\,325\,GHz to the atmospheric  water-vapour content the rms noise is $\sim$\,4 times higher and the flux calibration  could be compromised.

The emission at 322.61\,GHz, 322.33\,GHz, and 324.96\,GHz behaves similarly but with weaker and spatially somewhat more confined red-wing emission. Unfortunately, many of the \tioo lines are blended (both in frequency and spatially) with emission from other species. 

The discussion below is focussed on the morphology at 310.78\,GHz with Fig.~\ref{fig:TiO2_fullstructure} as a visual guide. Overall, the detected \tioo emission moves from west to east, across the stellar position, with velocities evolving from reddest to bluest. All lines show emission close to the star at the stellar \vlsr. In particular, all lines with $E_{\mathrm{up}}/k\gtrsim 360$\,K are centred on VY and emit mainly across $\vlsr\approx0-40$\,\kms (Fig.~\ref{fig:tio2all}), implying that these trace a central part of the outflow that may be accelerating.

At \underline{$\vlsr\geq44$\,\kms} the emission is mainly situated west of VY with a marginal contribution close to it. The emission is highly variable with $\vlsr$, doubly peaked in most velocity channels, and shows a hook-like feature at its western edge. The latter could relate to the north-west knot defined by \cite{humphreys2007_vycma_3dmorphology_kinematics} based on \emph{HST} observations, though more analysis is needed to investigate the nature of  this feature.

At \underline{$22<\vlsr<44$\,\kms} we find multiple emission peaks in most velocity channels, with those closest to VY slightly brighter than the west offset peaks. The integrated emission is elongated along a direction roughly parallel to the axis connecting the peak position of  C and VY, at a P.A. $\approx125^{\circ}$. 

At \underline{$18\leq\vlsr\leq22$\,\kms} the transitions at $\sim$310\,GHz show a bright tail extending south-west of VY at a P.A. $\approx220^{\circ}$, i.e. almost perpendicular to the axis connecting VY and C. This tail reaches $\sim1\arcsec$ away from VY, out to the south-west clump of \cite{shenoy2013_vycma_AO_2to5micron} and agrees very well with the features detected by e.g. \cite{smith2001_vycma} at wavelengths $\lambda\sim1-2.14$\,\um. The emission at 310.78\,GHz extends slightly further south-west than that at 310.55\,GHz. Since the most extended emission exceeds $5\sigma$, we suggest that the difference is real and that clumpiness in the outflow strongly influences the excitation of individual lines. 

At 312.73\,GHz and 312.82\,GHz we find signs of emission in the south-west tail, but these are likely artefacts from the cleaning procedure (see Sect.~\ref{sect:observations}). At 311.46\,GHz we see emission within the spatial region where the wouth-west tail is located, although at higher \vlsr than the transitions at $\sim$310\,GHz. This could be due to the south-west tail covering a wider velocity range than reported above, by a line blend, or a misidentification of the line as \tioo. We unfortunately do not have information for $\vlsr<22$\,\kms for this line.  No other transition shows detectable emission within the south-west tail, but most show a slight bulge of emission at $\sim$\,$0\farcs2$ south-west of the star, at the base of the south-west tail. Figure~\ref{fig:SW_all} shows a comparison for all lines to the south-west tail observed at 310.78\,GHz. 

\underline{At $\vlsr<18$\,\kms} the high-intensity emission is situated entirely east of VY, elongated, and oriented at $\sim$\,$15^{\circ}-25^{\circ}$ east from north. With bluer velocities the emission moves towards C and then appears to break up with a northern peak brighter than the southern one (e.g. Fig.~\ref{fig:310783}). Remarkable is that the low-intensity component of the 310.78\,GHz transition closely resembles the scattered light at 1\,\um. We find this strong correspondence at these blue velocities for no other \tioo transition. If the high- and low-intensity components have different intrinsic wind velocities, they could be spatially separated and trace different parts of the outflow.

\subsection{Excitation conditions}\label{sect:rotdiagram}
\begin{figure}
\centering
        \includegraphics[width=\linewidth]{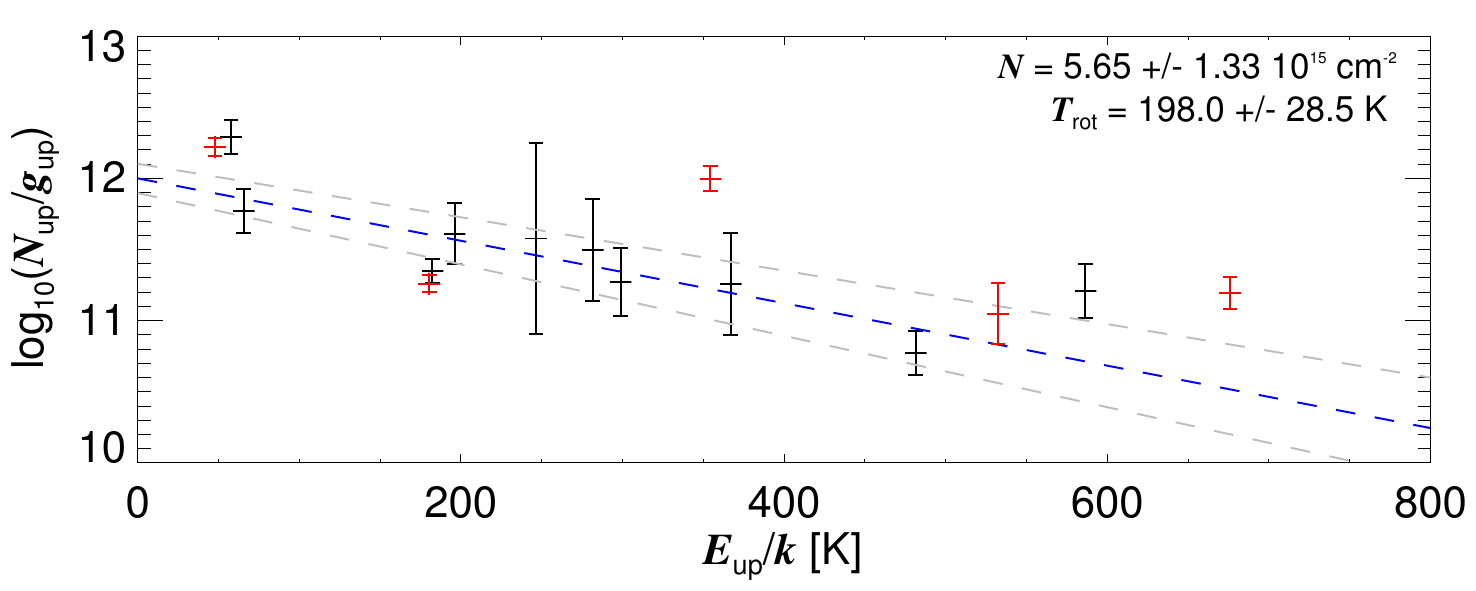}
        \caption{Rotational diagram. Lines indicated in red are blended or only partially covered in the observations and are excluded from the fitting procedure. All intensities, source sizes, and rms noise values are taken from Table~\ref{tbl:lineID}. The fit results and uncertainties are shown with the blue and grey dashed lines and are indicated at the top right.\label{fig:tio2_rot}}
\end{figure}

We investigate the excitation of \tioo lines via a rotational diagram analysis (Fig.~\ref{fig:tio2_rot}). Intensities, source sizes, and rms noise values are taken from Table~\ref{tbl:lineID}. Severely blended or only partially covered lines are excluded from the analysis. We derive a source-averaged column density $N_{\mathrm{col}}=5.65\pm1.33\times10^{15}$\,cm$^{-2}$ and a rotational temperature $T_{\mathrm{rot}}=198.0\pm28.5$\,K, in agreement with \citetalias{kaminski2013_tio_tio2}. We note that the kinetic temperature in the excitation region of \tioo varies from more than 1000\,K down to $\sim$100\,K \citep{decin2006_vycma}. Assuming an average mass-loss rate of $2\times10^{-4}$\,\msunyr \citep[e.g.][]{debeck2010_comdot}, an average velocity of 20\,\kms, and an 0\farcs9 diametric extent, at 1.2\,kpc, we find an abundance $\mathrm{TiO_2}/\mathrm{H_2}\approx3.8\pm0.9\times10^{-8}$.

The high dipole moment of \tioo \citep[6.33\,Debye;][]{wang2009_tio2} supports efficient radiative excitation and $T_{\mathrm{rot}}$ could hence reflect an average continuum brightness temperature as opposed to a gas kinetic temperature in the case of collisional excitation. Additionally, the large dipole moment induces electron-TiO$_2$ collision rates large enough to exceed H$_2$-TiO$_2$ collision rates if the fractional ionisation exceeds a few $10^{-6}$. We could not find a value for \vycma, but based on the result $n_{\mathrm{e}}/n_{\mathrm{H}}=3.8\times10^{-4}$ for the red supergiant \object{$\alpha$~Ori} \citep{harper2001_betelgeuse_extendedatmosphere} this type of excitation could be relevant\footnote{\vycma is of spectral type M2.5-M5e Ia \citep{houk1988_spectraltype}, $\alpha$~Ori of spectral type M1-2Ia-ab \citep{keenan1989_spectraltype}.}. We discuss the competition between collisional excitation and radiative excitation in more detail below,  in Sect.~\ref{sect:excitation}.


\section{Discussion}\label{sect:discussion}

\subsection{Excitation of titanium dioxide}\label{sect:excitation}
\begin{table*}[htb]
\caption{Summary of properties of selected transitions of \tioo. }\label{tbl:tio2_radtran}
\centering
\begin{tabular}{p{4cm}rrrrrr}
\hline\hline\\[-2ex]
$J'(K_a',K_c')$-$J''(K_a'',K_c'')$& \multicolumn{1}{c}{$\nu$} & \multicolumn{1}{c}{$A_{u,\ell}$} & \multicolumn{1}{c}{$\sum_{\ell} A_{u,\ell}$} & \multicolumn{1}{c}{$n_0$} & \multicolumn{1}{c}{$T_{\rm rad}$} & \multicolumn{1}{c}{$\rho_{\rm rad}$} \\
or vibrational band &  \multicolumn{1}{c}{(GHz)} & \multicolumn{1}{c}{(s$^{-1}$)} & \multicolumn{1}{c}{(s$^{-1}$)} & \multicolumn{1}{c}{(cm$^{-3}$)} & \multicolumn{1}{c}{(K)} & \multicolumn{1}{c}{(s$^{-1}$)} \\
\hline\\[-2ex]
$7(5,3)-6(4,2)$ & $312.248$ & $4.34\times10^{-3}$ & $5.04\times10^{-3}$ & $5\times10^7$ & $26.2$ &  $5.7\times10^{-3}$ \\
$7(5,3) - 7(4,4)$ & $206.094$ & $6.95\times 10^{-4}$ & \dots & \dots  & $15.3$ &  $7.6\times 10^{-4}$ \\
$7(5,3) - 8(4,4)$ & $84.567$ & $7.31\times 10^{-6}$ & \dots & \dots  & $7.15$ &  $9.6\times 10^{-6}$ \\
\\[-1ex]
$37(8,30)-37(7,31)$ & $324.966$ & $4.76\times10^{-3}$ & $4.63\times10^{-2}$ & $5\times 10^8$ & $27.7$ & $6.3\times10^{-3}$ \\
$37(8,30) - 36(9,27)$ & $205.259$ & $2.62\times 10^{-4}$ & \dots  &  \dots & $15.2$ &  $2.9\times 10^{-4}$ \\
$37(8,30) - 36(7,29)$ & $821.811$ & $2.96\times 10^{-2}$ & \dots  &  \dots & $94.3$ &  $5.7\times 10^{-2}$ \\
$37(8,30) - 36(5,31)$ & $1235.577$ & $1.17\times 10^{-2}$ & \dots  &  \dots & $132.4$ &  $2.0\times 10^{-2}$ \\
\\[-1ex]
$\nu_1=1-0$ & $28840$ & $3.6$ & $3.6$ & \tablefootmark{a} & $350$ & $7.2\times 10^{-2}$ \\
$\nu_2=1-0$ & $9863$ & $7.2\times 10^{-2}$ & $7.2\times 10^{-2}$ &  \tablefootmark{a} & 300 & $2.0\times 10^{-2}$ \\
$\nu_3=1-0$ & $28030$ & $42$ & $42$ &  \tablefootmark{a} & 350 & $9.0\times 10^{-1}$ \\
\hline
\end{tabular}
        \tablefoot{Columns list the transitions (rotational and vibrational), the frequency $\nu$ at which these transitions occur, their transition probability  $A_{u,\ell}$,  inverse lifetime  $\sum_{\ell} A_{u,\ell}$, critical density $n_0$, radiative temperature $T_{\rm rad}$ and the calculated pumping rate $\rho_{\rm rad}$. 
\tablefoottext{a}{Appropriate collision rates are unknown; therefore, no critical density is tabulated for the vibrational transitions.}
        }
\end{table*}

Many effects need to be taken into account in a full treatment of the excitation and radiative transfer of rotational transitions of TiO$_2$. Important molecular data are lacking: although the rotational energy levels and radiative data are well determined for TiO$_2$ in CDMS \citep{mueller2001_cdms,mueller2005_cdms,bruenken2008_tio2},  no rates exist for excitation of TiO$_2$ by hydrogen-impact nor have the vibration-rotation spectra been fully analysed. 

For illustration of the transition rates for \tioo, consider two of the observed rotational transitions at 312.248 and 324.965\,GHz with properties summarised in Table~\ref{tbl:tio2_radtran}. Collisional excitation at the kinetic temperature of the gas can dominate only when the downward rate of collision-induced transitions greatly exceeds the downward rates of radiative transitions for each upper state. This condition can be translated to a critical density $n_0 = \sum_{\ell} A_{u,\ell} / q_0$, where $A_{u,\ell}$ is the spontaneous transition probability of each transition from upper state $u$ to lower state $\ell$. To estimate this density, we assume a characteristic\footnote{Such a value is typical of the largest downward collisional rates for H$_2$ collisions with a heavy, polar molecule. For example, accurate collision rates have been computed by \cite{cernicharo2011_so2_collisionalexcitation} for H$_2$ on SO$_2$, a heavy molecule  with a relatively large dipole moment (1.63\,Debye). The SO$_2$ quenching rates at low temperature are $\sim 2\times 10^{-10}$.} quenching rate coefficient $q_0=10^{-10}$ cm$^3$ s$^{-1}$ for rotational transitions induced by collisions with neutral species H or H$_2$. Values for $n_0$ are  listed in Table~\ref{tbl:tio2_radtran}. Assuming a mass-loss rate of $10^{-4}$\,\msunyr, such densities are however reached only at distances from the star lower than $\sim$0\farcs065, whereas \tioo is excited over a much larger region, where the density quickly decreases to a few $10^5$\,cm$^{-3}$.

In addition, collision rates for electron impact have been computed in the Born approximation. Because the electric dipole moment of \tioo is so large, electron-impact rates are expected to show a strong propensity for radiatively allowed transitions. Computed rates are likely to be accurate within 50\%. Under an assumed very low fractional electron density of $10^{-7}$, the collision-induced downward transition rate for the 312\,GHz transition is for  5\% due to electron collisions.  Assuming the fractional ionisation reported for $\alpha$\,Ori ($3.8\times10^{-4}$) electron collisions would completely dominate over neutral (hydrogen) collisions by a factor of the order of 100 or more and might compete well with infrared pumping.

We note that the upper state of the 312\,GHz transition decays mainly by the observed transition itself, while the upper state of the high-excitation 324\,GHz transition is depopulated more rapidly by submm-wave transitions at 821 and 1235\,GHz. As a consequence, a much higher density would be required to excite the 324\,GHz line by collisions than the 312\,GHz line.

The continuum intensity of VY CMa is so large at infrared and submm wavelengths that radiative excitation (pumping) must be taken into account. If the observed continuum flux of VY CMa is assumed to arise within a $0\farcs3$ diameter region \citepalias{ogorman2015_alma_vycma} and to be diluted by a geometrical factor $1/9$ over the $0\farcs9$ extent of the observed TiO$_2$ emission, then the average Planckian radiation brightness temperatures are estimated to be $T_{\rm rad} \geq 25$ K at 312 to 324\,GHz. The corresponding pumping rates (absorption and stimulated emission) in this radiation field can be expressed as 
$$ \rho_{\rm rad} = \frac{A_{u,\ell}}{\exp\bigl(h\nu/kT_{\rm rad}\bigr) - 1}$$
for each radiative transition that connects states of interest. The basic radiative data for TiO$_2$ are collected in Table~\ref{tbl:tio2_radtran}. Although the fundamental bands of the three vibrational modes have not been fully  analysed rotationally, the band frequencies and band strengths are approximately known \citep{grein2007}. In the adopted continuum model, the infrared intensity is high enough to drive absorption in vibration-rotation lines at rates of the order of 0.01 to 1.0 s$^{-1}$. In order for collisional excitation in a pure rotational transition $u\to \ell$ to compete with infrared pumping in VY CMa, a density much greater than  
$$  n_0 \approx  \Bigl(  \rho(\nu_1) + \rho(\nu_2) + \rho(\nu_3) + \sum_{\ell} A_{u,\ell}
\Bigr) / q_0 \approx  10^{10} \;\;\;{\rm cm}^{-3} $$
would be required. The radiative rates in Table~\ref{tbl:tio2_radtran} suggest that radiative excitation is likely to be very important for TiO$_2$ in VY CMa. Therefore, the rotational temperature derived in Sect.~\ref{sect:rotdiagram} might have no direct relationship to the kinetic temperature. Because the observed transitions in the present study span a wide range of excitation energies, the simple rotation-diagram analysis still provides a useful first estimate of the molecular column density and abundance. 

\subsection{Outflow components}\label{sect:outflows}
From the comparison to the continuum at $\sim$\,321\,GHz, we derive that \tioo is excited in the directions with lower dust densities. The absence of detected \tioo emission north of the star could then point to efficient obscuration of this part of the outflow, in line with the observations of e.g. \cite{smith2001_vycma}. Attenuation of the stellar radiation field to the east and west of the star is limited. From this, we do not expect an equatorial enhancement of the mass-loss rate, since this would likely have induced more efficient dust formation, which is not seen in the continuum. 

The \tioo emission traces multiple wind components. We find a red outflow to the west and a blue outflow to the east. With the clear exception of the interaction of the \tioo gas with clump C in the east, the two seem roughly symmetric around the star and aligned with the axis connecting VY and C. We rule out an equatorially enhanced environment such as an expanding disk or ring, based on the spatial distribution of the \tioo emission at different \vlsr. We rather suggest an accelerating bipolar-like outflow at lower densities. We also find a predominantly blue south-west outflow, connecting the star and the south-west clump, approximately perpendicular to the VY\,--\,C connecting axis. We find no north-east counterpart in \tioo emission, likely implying that the south-west tail is indeed caused by an event in one preferred direction, as opposed to a bipolar event.

\subsection{Interaction of blue outflow with clump C}\label{sect:blue}
Whereas the H$_2$O maser emission in the ``valley'' between VY and C implies that C is close to or in the plane of the sky \citepalias{richards2014_alma_vycma}, the observations suggest that the \tioo gas breaks up around C while moving towards the observer, placing C -- at least partially -- in front of the plane of the sky. We therefore deem it likely that the H$_2$O masers and the \tioo emission probe parts of the outflow east of VY with different physical properties. Whereas the masers are probably excited through shocks at high densities, \tioo is more likely excited through radiation, at lower outflow densities. In the denser regions, \tioo might not be excited and/or it might be efficiently depleted from the gas phase. The latter is, however, less likely (see below). We therefore suggest that \tioo traces the blue-shifted wind to the east of VY with lower densities which runs into and curves around C.

\subsection{Titanium dioxide and scattered light}\label{sect:swtail}
\citetalias{kaminski2013_vycma_sma} reported emission offset by $\sim$\,$1\arcsec$ south-west from the central molecular emission for multiple species. However, owing to lower sensitivity they did not find indications for this in the TiO$_2$ emission, whereas we clearly detect the south-west tail at $\sim$310\,GHz (see Sect.~\ref{sect:morphology}). 

The likelihood of radiative excitation of \tioo and the agreement between the south-west tail in the \tioo emission and the scattered stellar light at 1\,\um from \cite{smith2001_vycma} suggest the presence of a tail of gas and dust lit up by a stellar radiation field less attenuated than at other position angles. The bulge south-west of the star seen for most \tioo transitions is consistent with this. We infer a lower degree of attenuation and from that a lower (dust) density south-west of the star. We speculate that the suggested localised ejection leading to the formation of the south-west clump \citep{shenoy2013_vycma_AO_2to5micron} could have created a cavity in the dense material close to the star. 

Given the agreement with the blueshifted \tioo emission, the outflow traced by the scattered light is likely oriented out of the plane of the sky. The lack of consistent \vlsr coverage over the different \tioo transitions, however, further complicates the three-dimensional and kinematic constraints. In-depth analysis of all other emission lines detected with ALMA is needed to constrain the basic properties of the south-west tail.

\subsection{Titanium dioxide and dust formation}
If the nucleation of \tioo can occur at $\lesssim$\,2000\,K \citep[e.g.][]{lee2014_tio2_sio_nucleation}, one expects \tioo depletion from the gas phase initiated close to the star, i.e. within a few stellar radii. From the ALMA observations \citetalias{ogorman2015_alma_vycma} put an upper limit to the dust-condensation radius of $\sim0\farcs06\pm0\farcs02\approx$\,10 stellar radii. The \tioo emission is present close to the star and out to $\lesssim$0\farcs45 in  directions without extended continuum emission. Where \tioo and the bulk dust coexist, the emission extends to $\sim$0\farcs20, well beyond the dust-condensation radius. Since the \tioo emission is seen at radial distances from the star far beyond the dust-condensation region, and along directions where dust continuum is observed, we claim that \tioo is not a tracer for low grain-formation efficiency. The strong correspondence between the \tioo emission and the south-west tail of dust-scattered stellar light (see Sect.\ref{sect:morphology} and Fig.~\ref{fig:TiO2_fullstructure}) supports this. We also considered the high derived column density and suggest that \tioo plays only a minor role as a primary dust seed around \vycma. 

Since the ALMA observations covered no transitions of TiO, we cannot expand the discussion of \citetalias{kaminski2013_tio_tio2} on the relation between TiO, \tioo, and the dust. 


\section{Conclusion}\label{sect:conclusion}
We detect 15 transitions of \tioo in high-resolution ALMA observations of the red supergiant \vycma. The main emission region spans $\sim$0\farcs9 ($\sim$1080\,AU) in a roughly east-west oriented direction, centred on the star. The behaviour of the gas is complex throughout \vlsr-space with bright peaks appearing only in very confined \vlsr-ranges, implying that the outflow is very clumpy, on spatial scales not resolved by the current observations with a $0\farcs23\times0\farcs13$ beam. We observe a tail of \tioo emission extending out to $\sim$1\arcsec\ south-west of the star, consistent with structures seen in the optical and near-infrared. It is oriented out of the plane of the sky and mainly covers projected velocities of a few \kms, but reaches up to $\sim$40\,\kms in some cases. We suggest that the \tioo in this tail is illuminated by stellar radiation penetrating through a low-density cavity in the south-west part of the circumstellar environment of the star, potentially created by the ejection that led to the south-west clump. Within a bipolar-like \tioo outflow, the blue-shifted emission exhibits a strongly different orientation and behaviour than the red-shifted emission, suggesting that the stellar wind runs into the large body of dust situated $\sim$0\farcs335 ($\sim$400\,AU) south-east of the star. 

We suggest that \tioo might play only a minor role in the dust-condensation process in the complex outflow around \vycma, and potentially also around other oxygen-rich evolved stars with extreme mass outflows. High-resolution imaging is however still needed to correlate the emission of \tioo with that of TiO, and to further investigate the relative importance of silicon, titanium, and other metals in the dust condensation. 


\begin{acknowledgements}
This paper makes use of the following ALMA data: ADS/JAO.ALMA2011.0.00011.SV. ALMA is a partnership of ESO (representing its member states), NSF (USA) and NINS (Japan), together with NRC (Canada) and NSC and ASIAA (Taiwan), in cooperation with the Republic of Chile. The Joint ALMA Observatory is operated by ESO, AUI/NRAO and NAOJ. WV and EOG acknowledge support from the ERC through consolidator grant 614264. MM has received funding from the People Programme (Marie Curie Actions) of the EU’s FP7 (FP7/2007-2013) under REA grant agreement No. 623898.11.
\end{acknowledgements}


\addcontentsline{toc}{chapter}{Bibliography}
\bibliographystyle{aa}
\bibliography{25990.bib}


\Online
\begin{appendix}

\section{Maps of titanium dioxide emission}\label{app:channels}
\begin{figure*}[p]
\includegraphics[width=\linewidth]{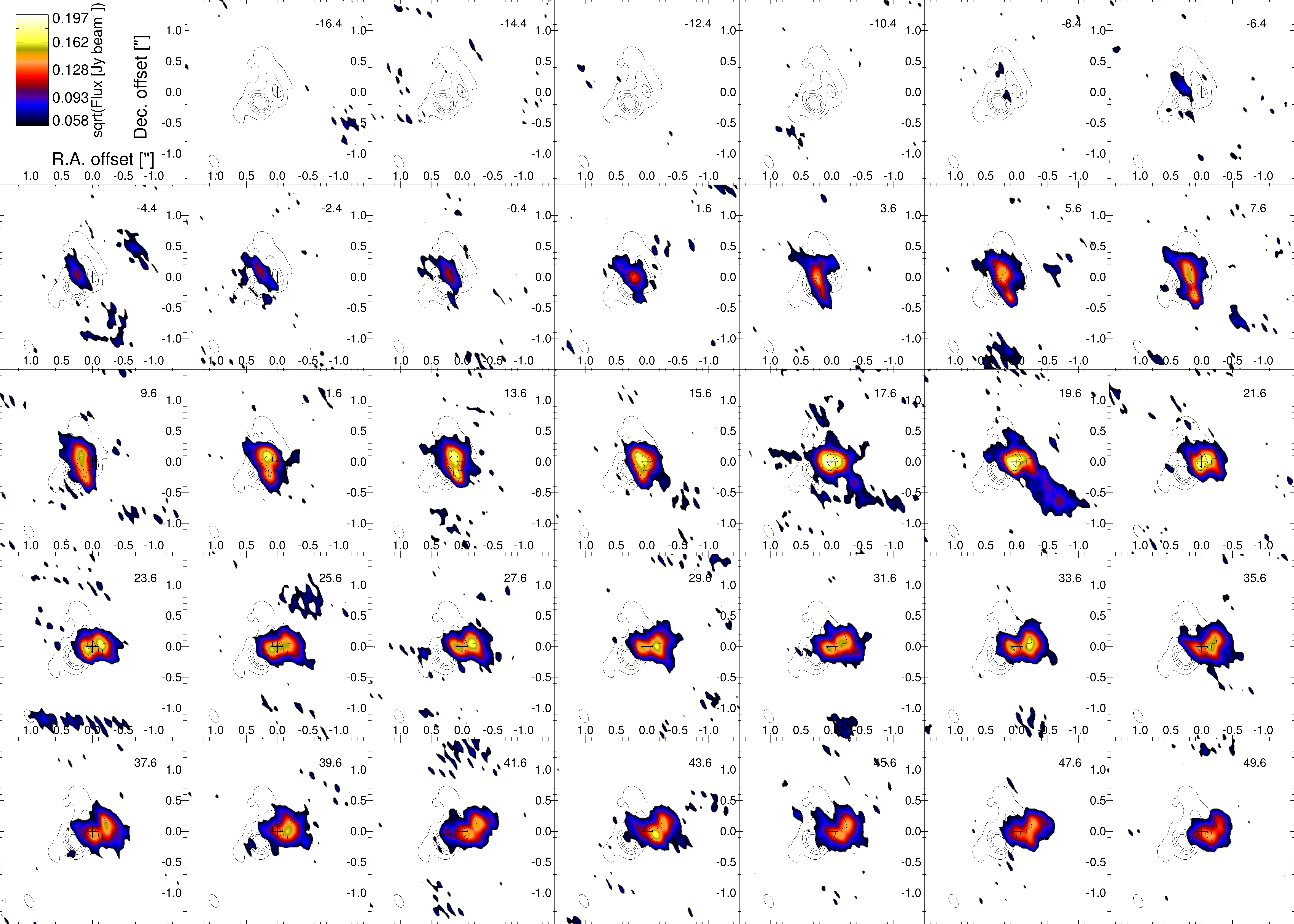}
\caption{Channel maps of the TiO$_2$ emission at 310.55\,GHz, at a 2\,\kms velocity resolution. Black contours show the continuum measured with ALMA at 321\,GHz \citepalias{ogorman2015_alma_vycma,richards2014_alma_vycma}. The stellar position is indicated with a white cross. Spatial scales are indicated in the top left panel and are the same for all panels. The colour scale starts at the $3\sigma$ level and is plotted as the square root of the flux for increased contrast. }\label{fig:310554}
\end{figure*}

\begin{figure*}[p]
\centering
\includegraphics[width=.9\linewidth]{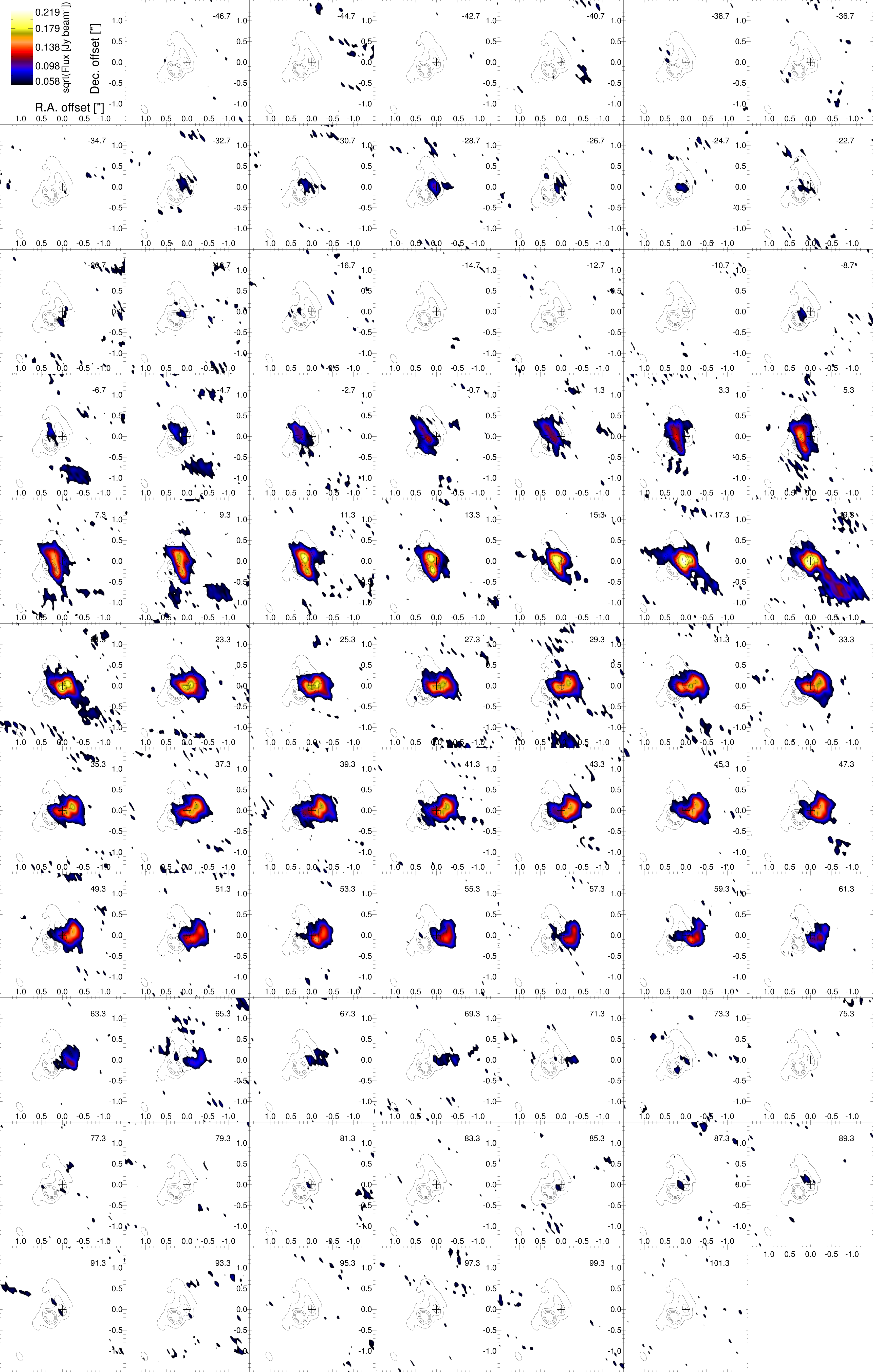}
\caption{Same as Fig.~\ref{fig:310554}, but for TiO$_2$ emission at 310.78\,GHz.}\label{fig:310783}
\end{figure*}

\begin{figure*}[p]
\includegraphics[width=\linewidth]{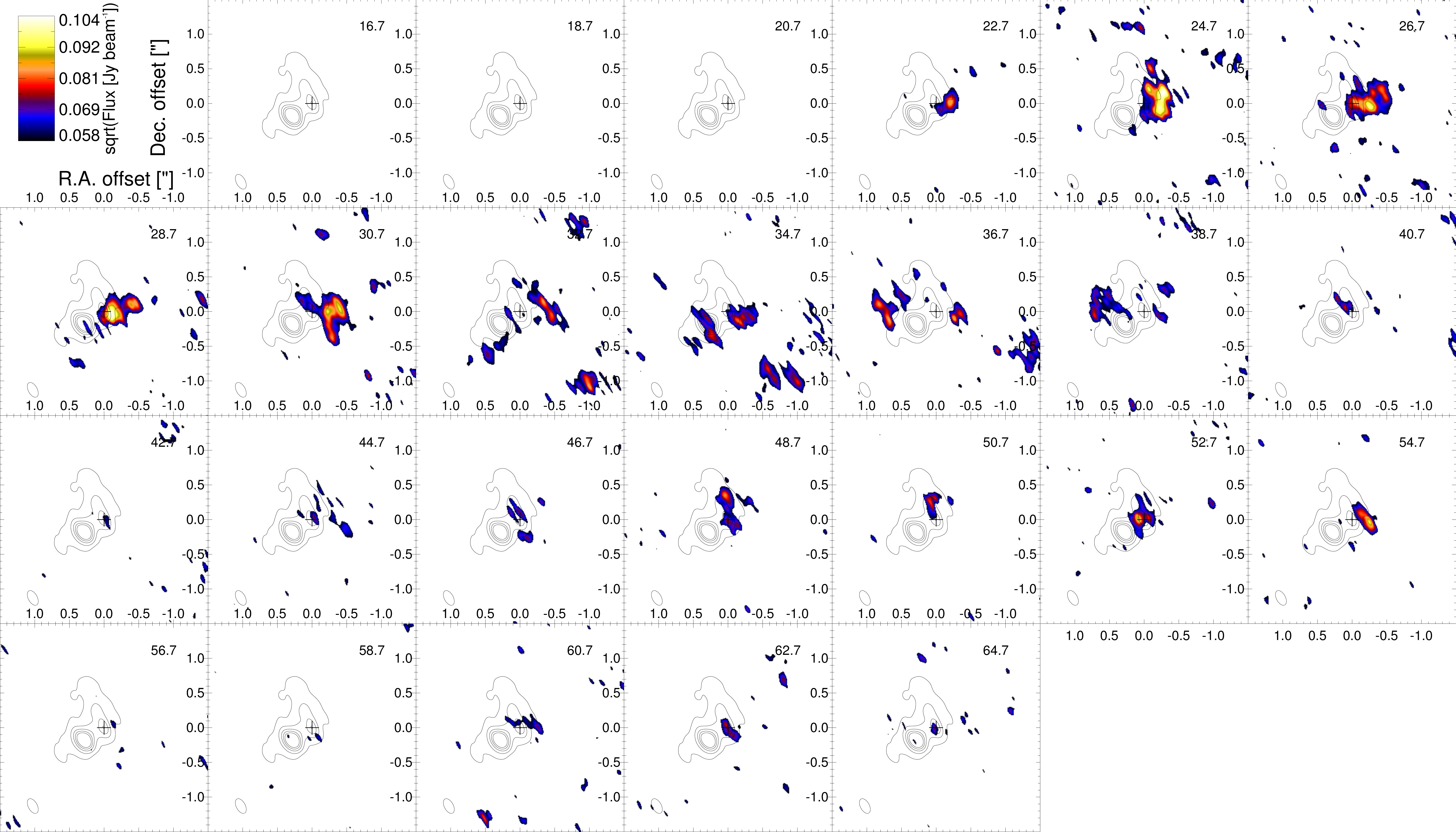}
\caption{Same as Fig.~\ref{fig:310554}, but for TiO$_2$ emission at 311.46\,GHz.}\label{fig:311462}
\end{figure*}

\begin{figure*}[p]
\centering
\includegraphics[width=.9\linewidth]{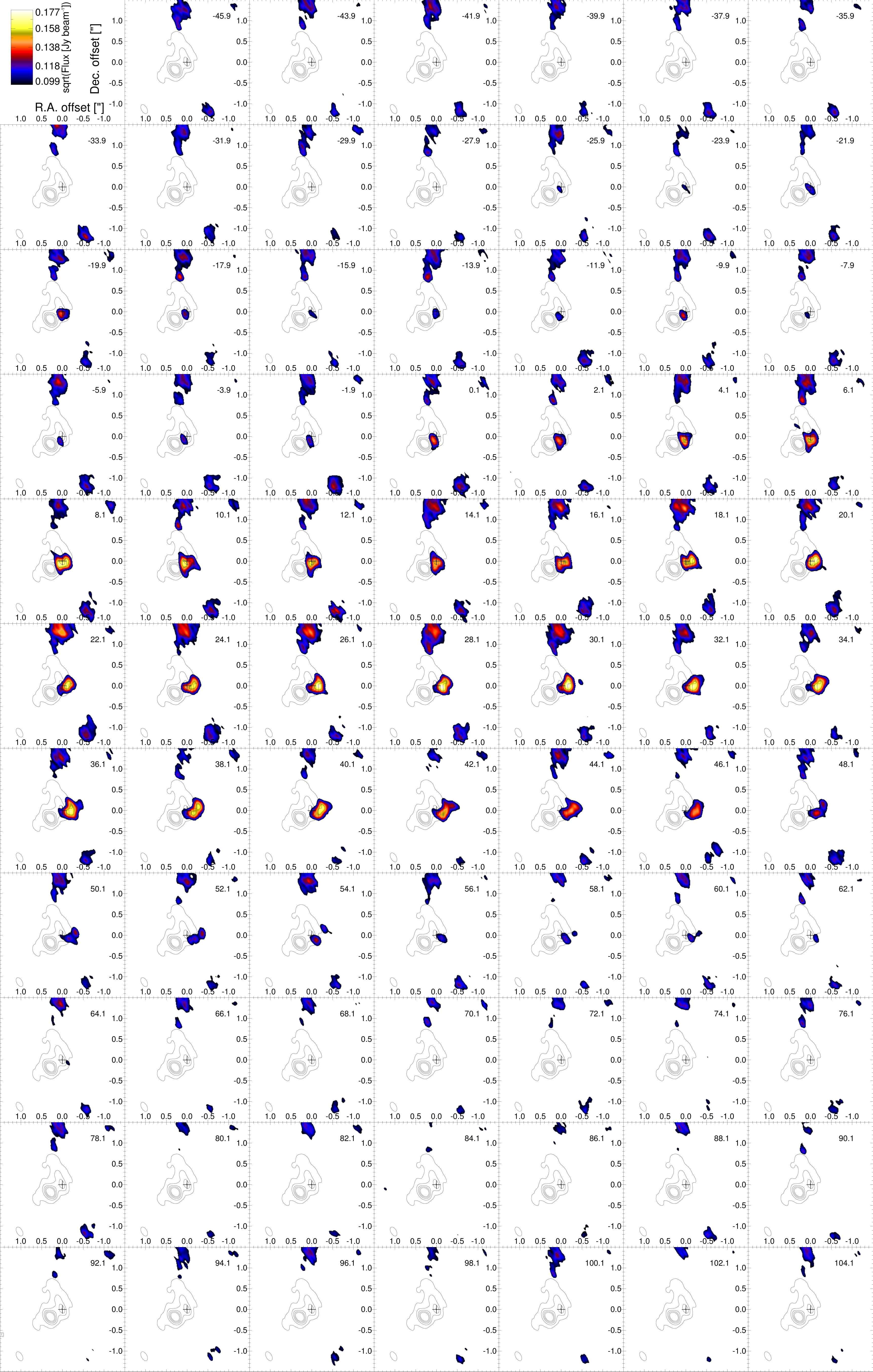}
\caption{Same as Fig.~\ref{fig:310554}, but for TiO$_2$ emission at 312.25\,GHz. }\label{fig:312248}
\end{figure*}

\begin{figure*}[p]
\includegraphics[width=\linewidth]{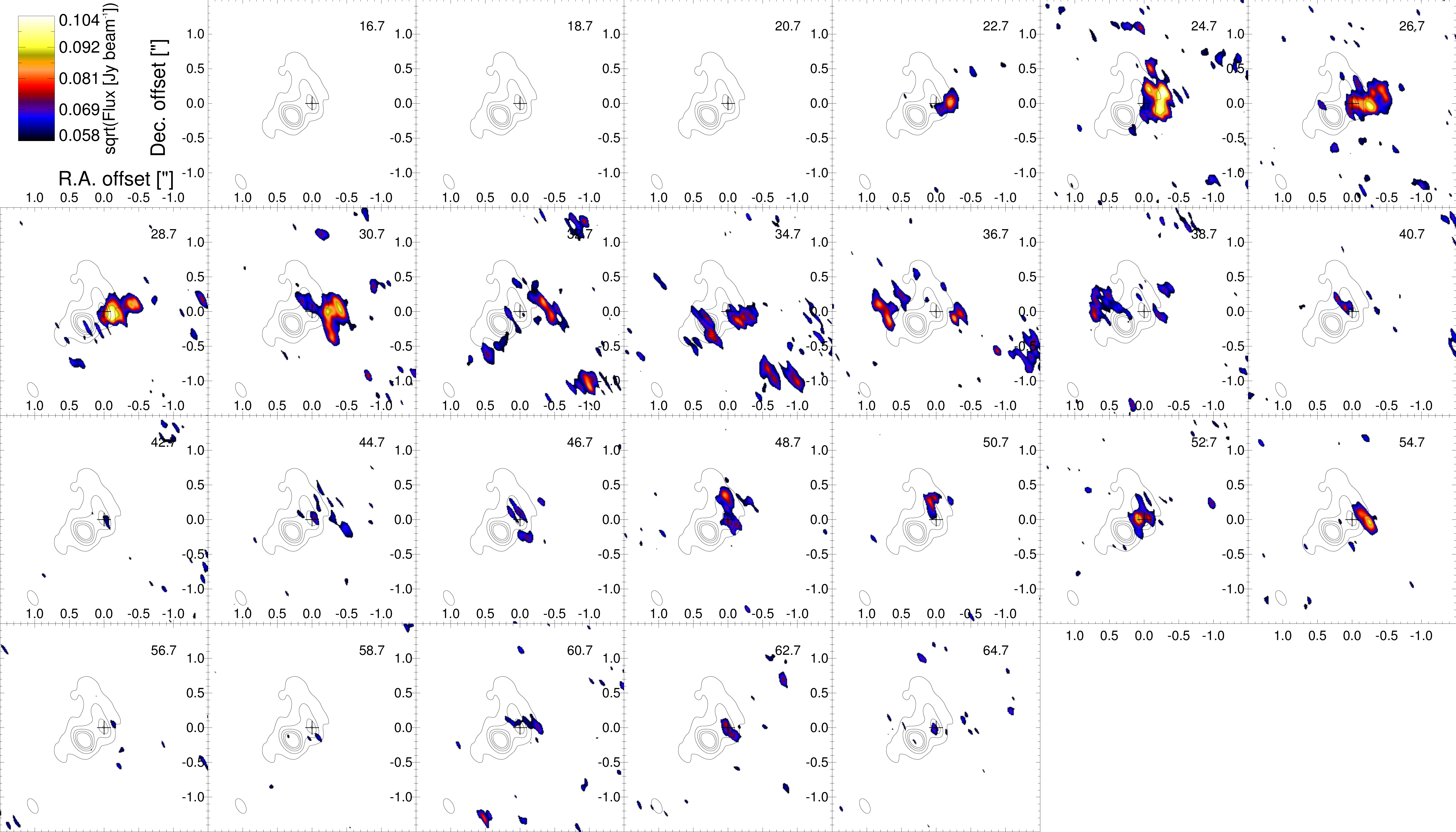}
\caption{Same as Fig.~\ref{fig:310554}, but for TiO$_2$ emission at 312.73\,GHz. }\label{fig:312732}
\end{figure*}

\begin{figure*}[p]
\includegraphics[width=\linewidth]{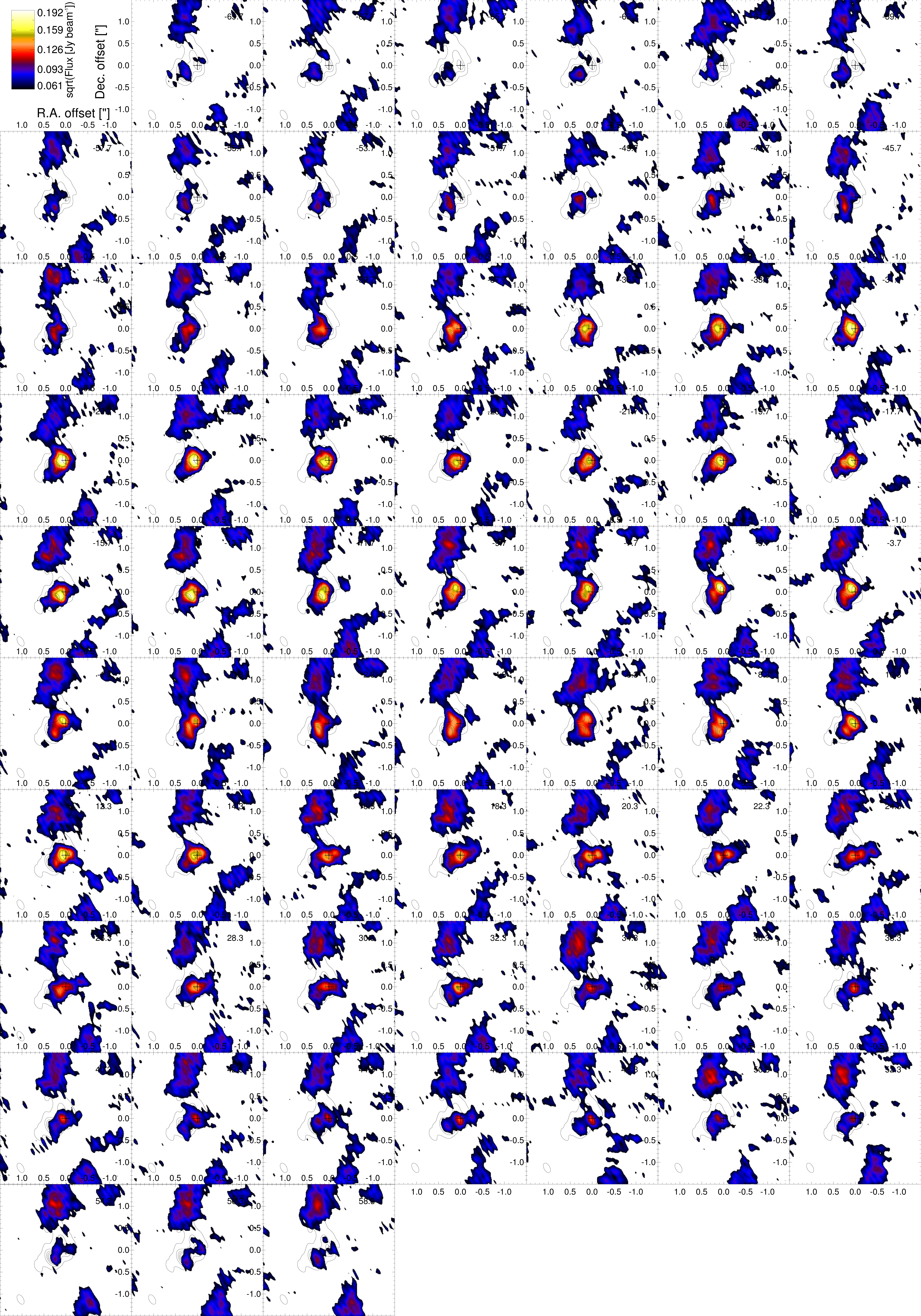}
\caption{Same as Fig.~\ref{fig:310554}, but for TiO$_2$ emission at 312.82\,GHz. }\label{fig:312817}
\end{figure*}

\begin{figure*}[p]
\includegraphics[width=\linewidth]{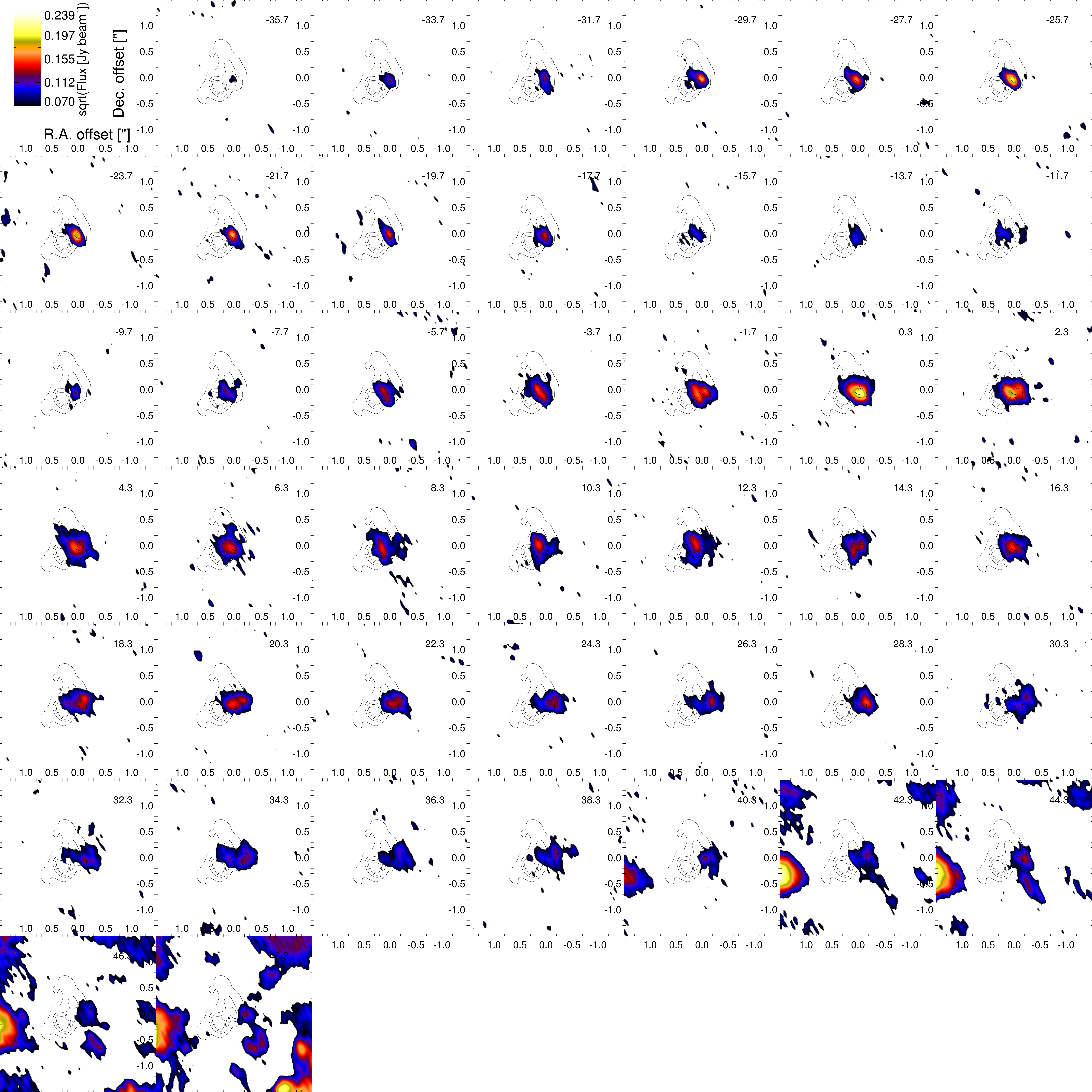}
\caption{Same as Fig.~\ref{fig:310554}, but for TiO$_2$ emission at 321.40\,GHz. Artefacts in channels with $\vlsr\geq55$\,\kms are due to the presence of strong SO$_2$ emission. See also Fig.~\ref{fig:tio2all}.}\label{fig:321402}
\end{figure*}

\begin{figure*}[p]
\includegraphics[width=\linewidth]{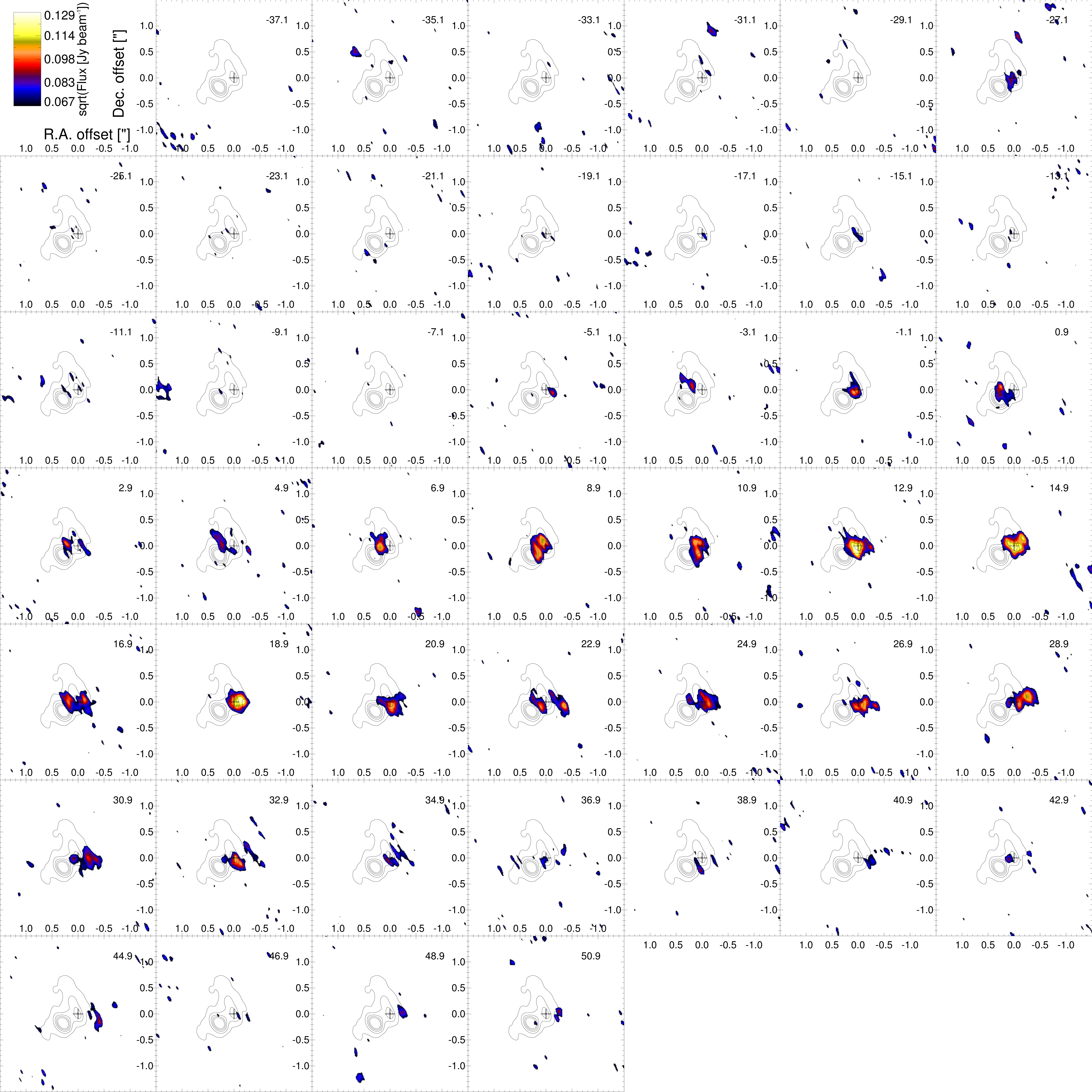}
\caption{Same as Fig.~\ref{fig:310554}, but for TiO$_2$ emission at 321.50\,GHz.}\label{fig:321501}
\end{figure*}

\begin{figure*}[p]
\centering
\includegraphics[width=.9\linewidth]{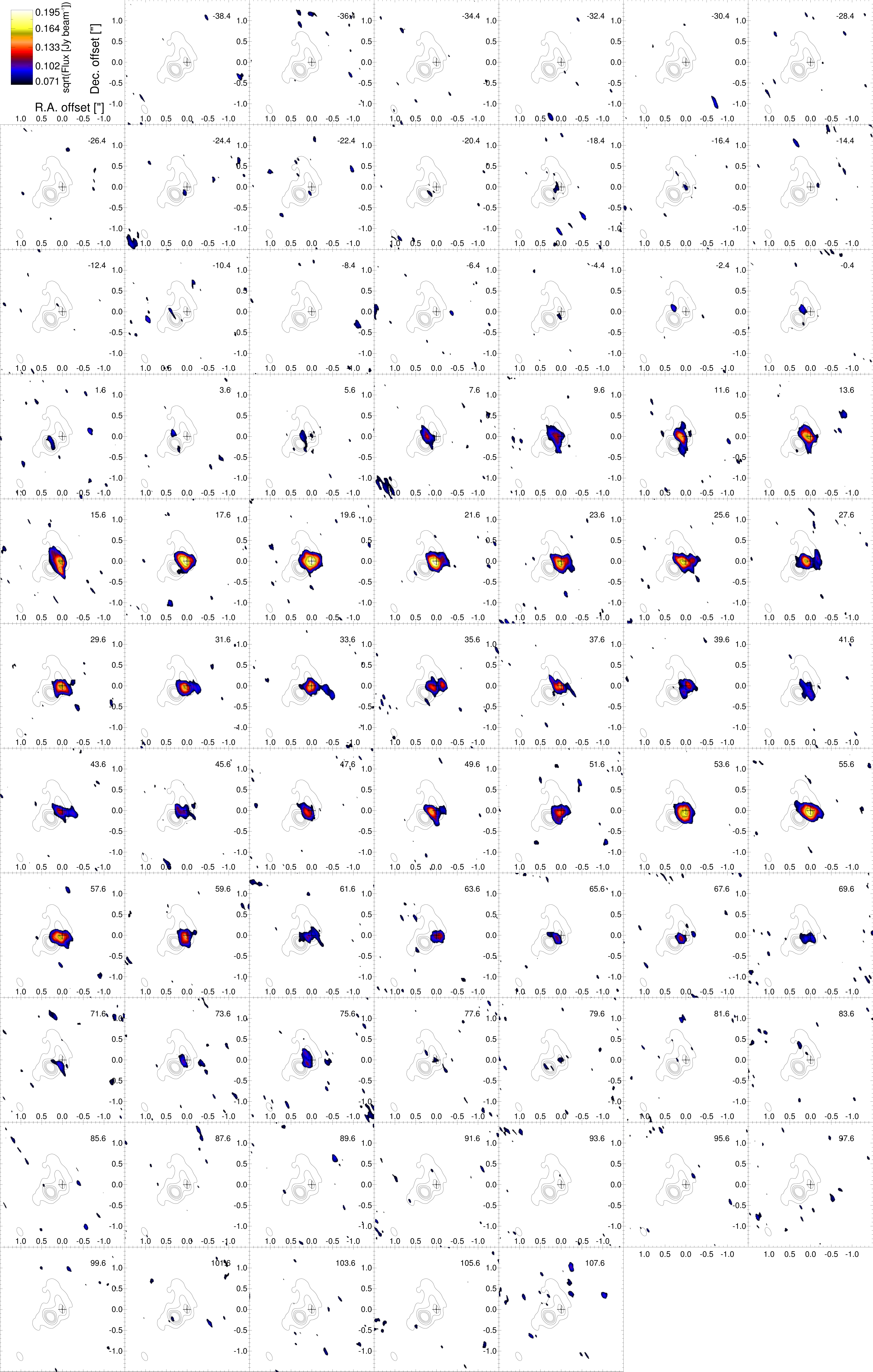}
\caption{Same as Fig.~\ref{fig:310554}, but for TiO$_2$ emission at 322.33\,GHz.}\label{fig:322334}
\end{figure*}

\begin{figure*}[p]
\includegraphics[width=\linewidth]{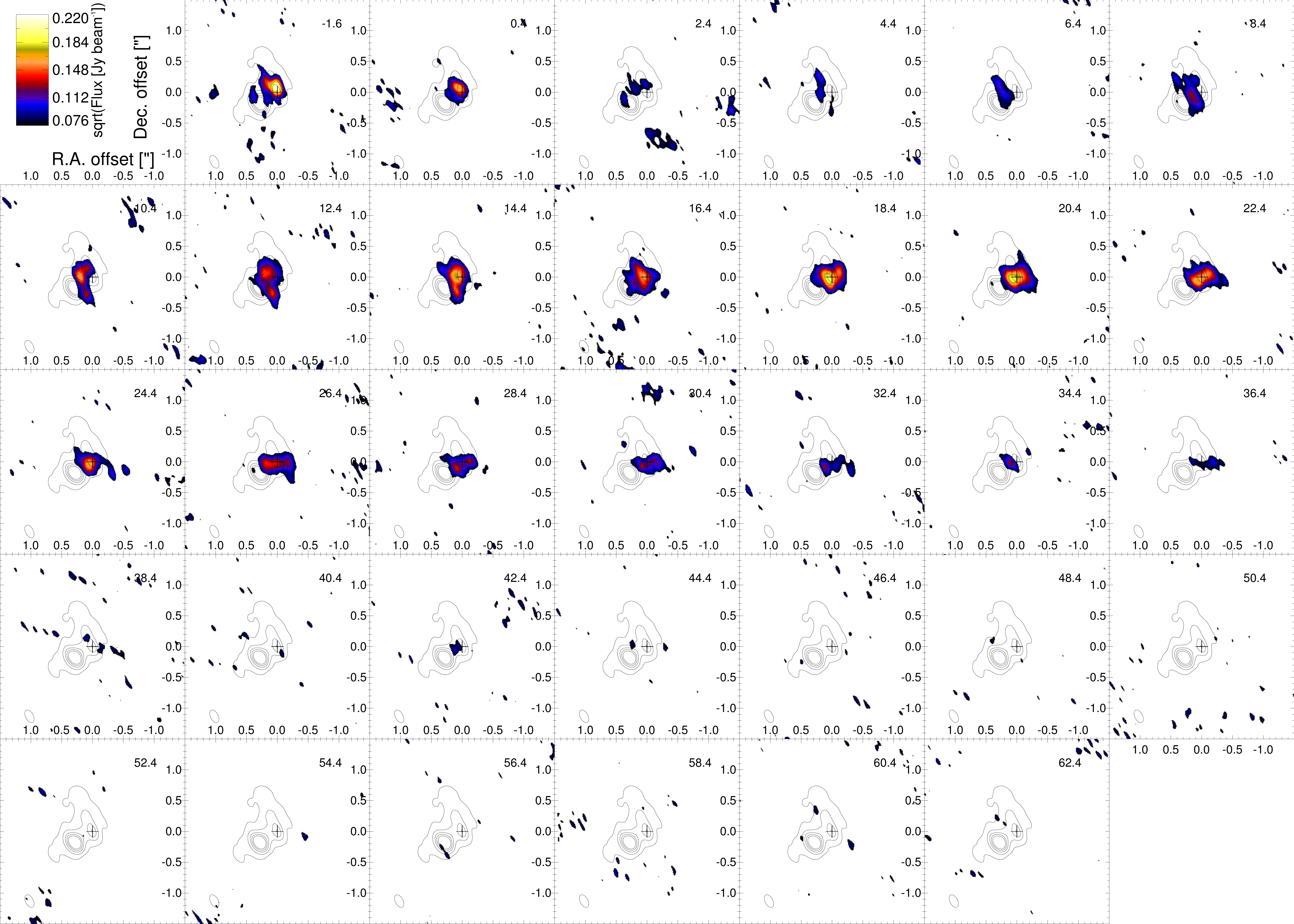}
\caption{Same as Fig.~\ref{fig:310554}, but for TiO$_2$ emission at 322.61\,GHz.}\label{fig:322613}
\end{figure*}

\begin{figure*}[p]
\includegraphics[width=\linewidth]{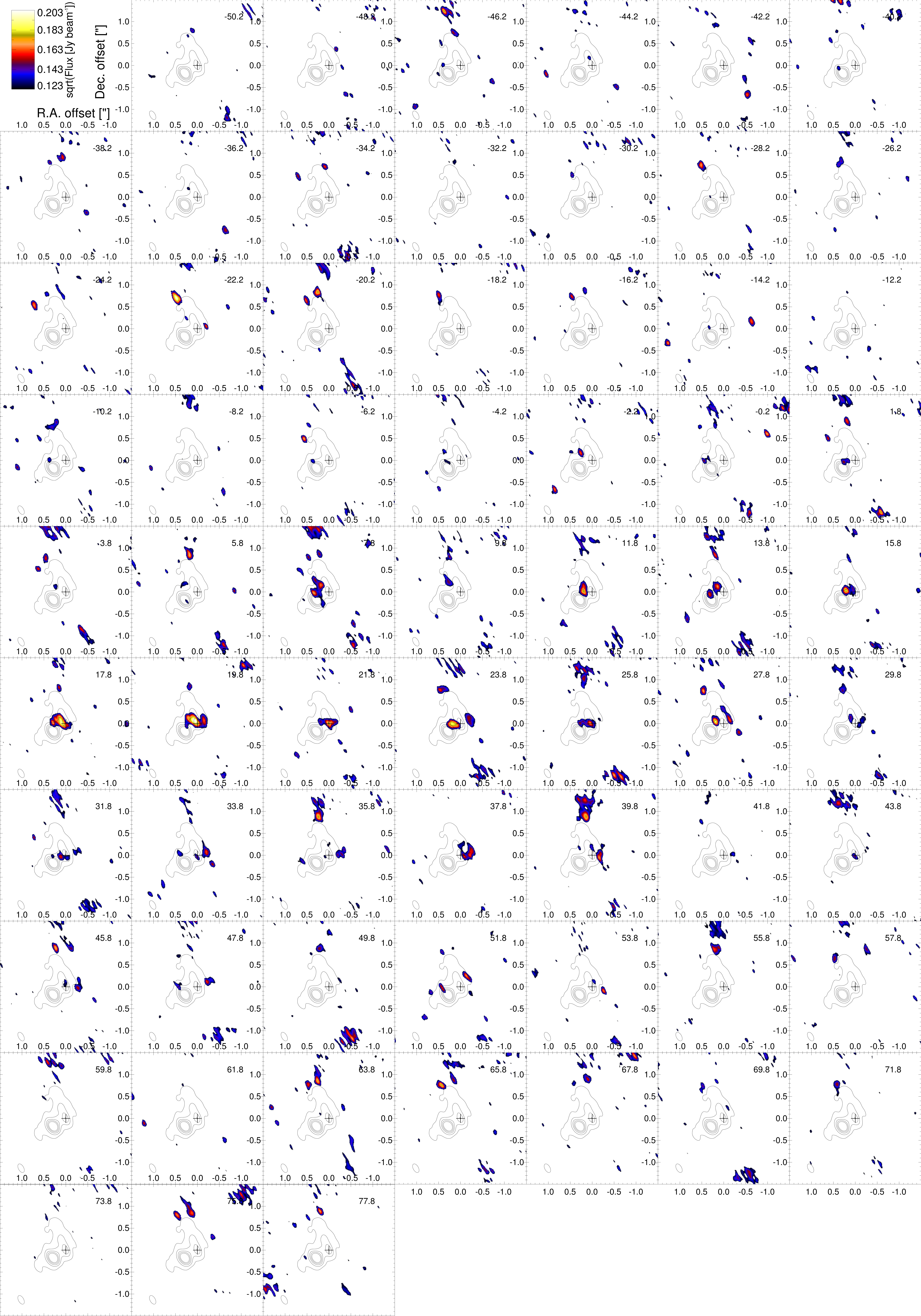}
\caption{Same as Fig.~\ref{fig:310554}, but for TiO$_2$ emission at 324.49\,GHz.}\label{fig:324493}
\end{figure*}

\begin{figure*}[p]
\includegraphics[width=\linewidth]{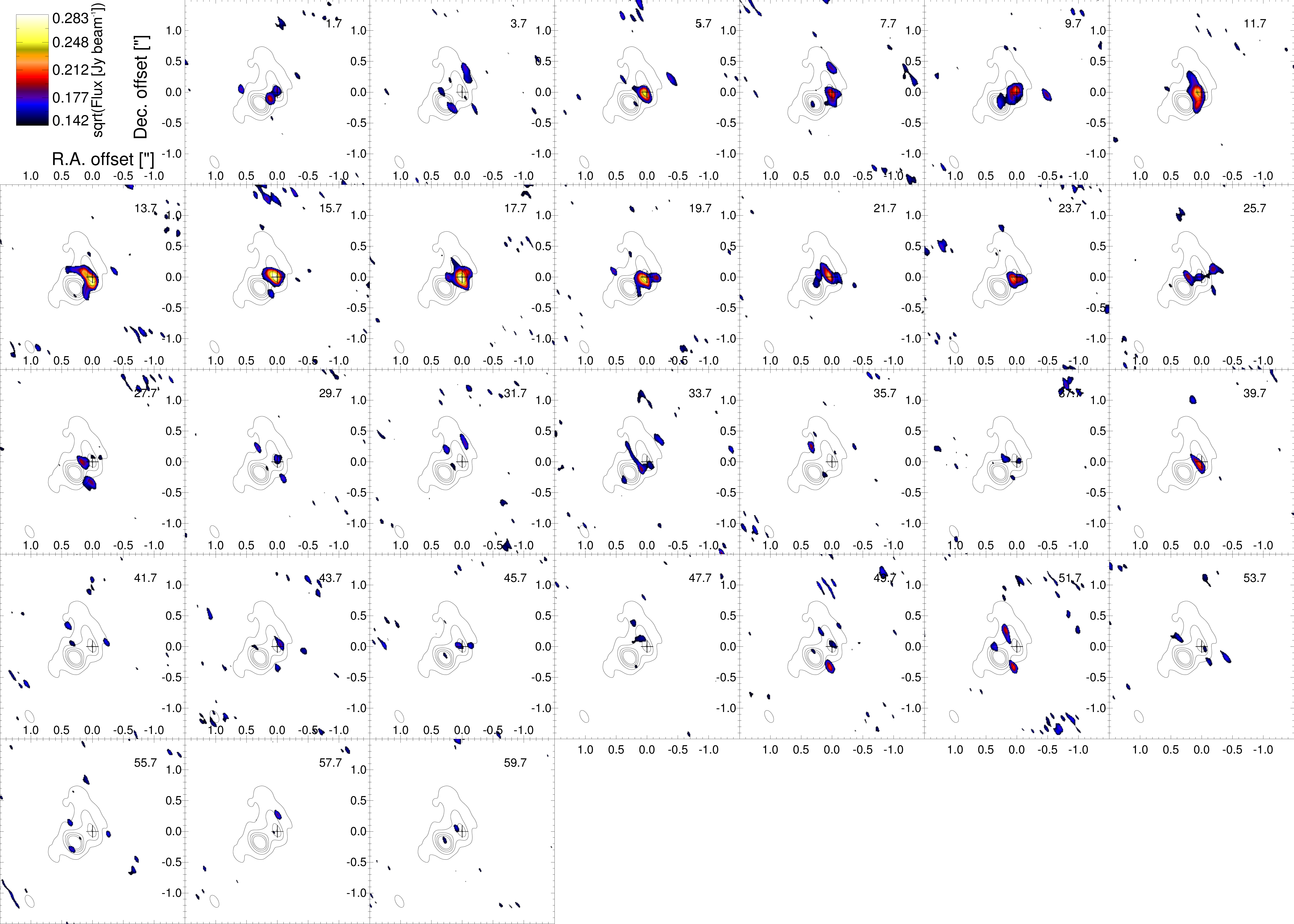}
\caption{Same as Fig.~\ref{fig:310554}, but for TiO$_2$ emission at 324.96\,GHz.}\label{fig:324966}
\end{figure*}

\begin{figure*}[p]
\includegraphics[width=\linewidth]{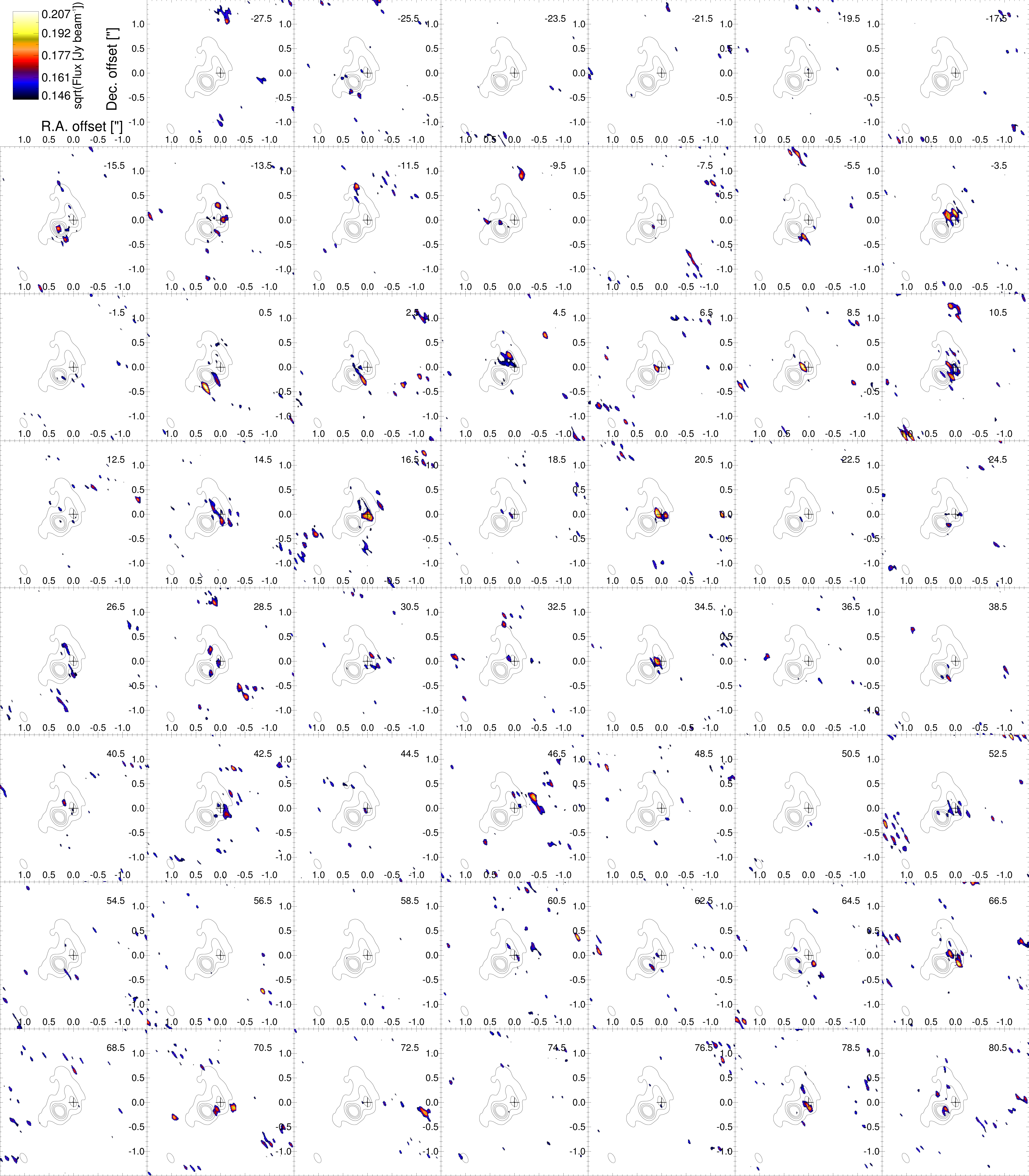}
\caption{Same as Fig.~\ref{fig:310554}, but for TiO$_2$ emission at 325.32\,GHz.}\label{fig:325322}
\end{figure*}

\begin{figure*}[p]
\includegraphics[width=\linewidth]{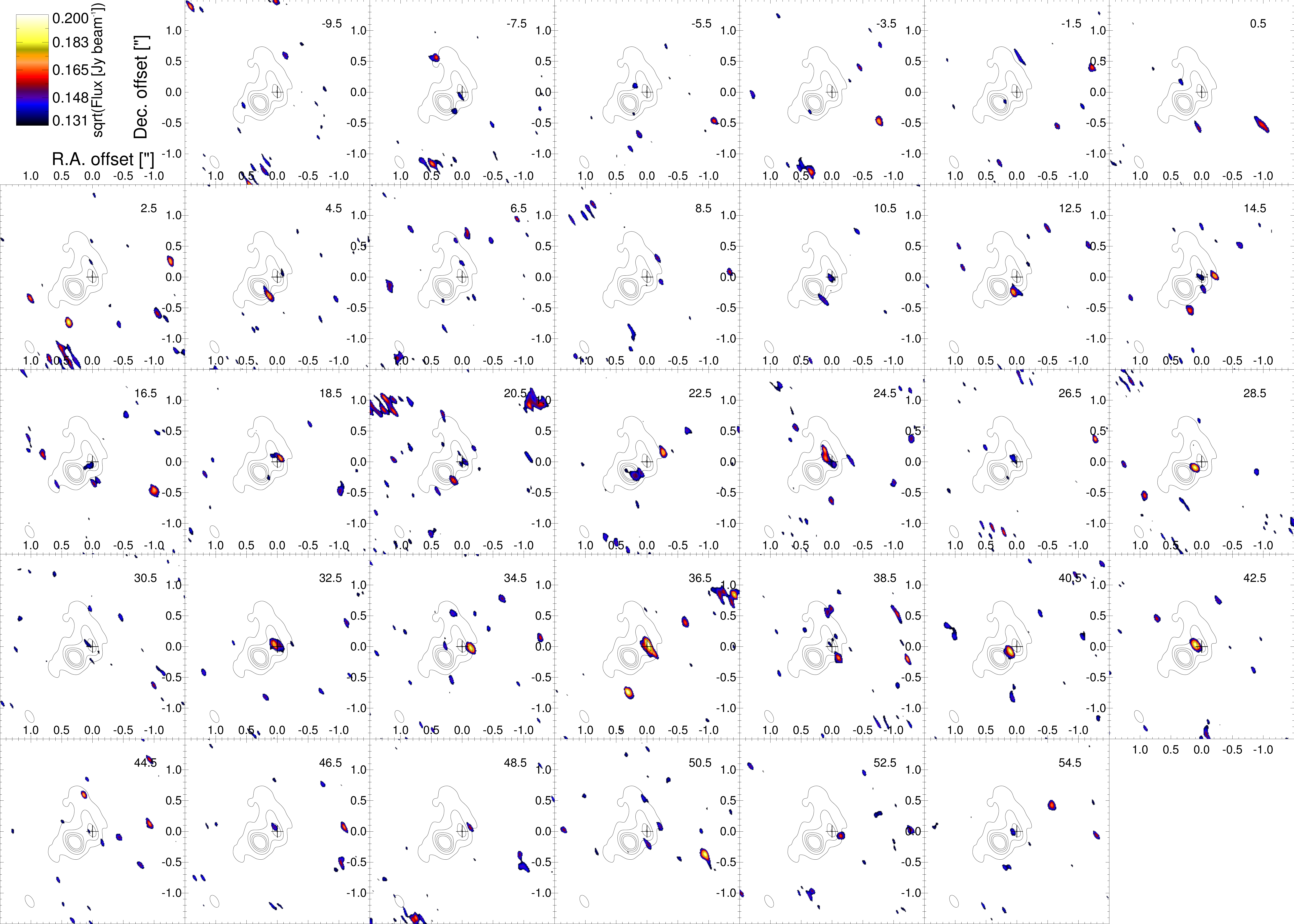}
\caption{Same as Fig.~\ref{fig:310554}, but for TiO$_2$ emission at 325.50\,GHz.}\label{fig:325500}
\end{figure*}

\begin{figure*}[p]
\includegraphics[width=\linewidth]{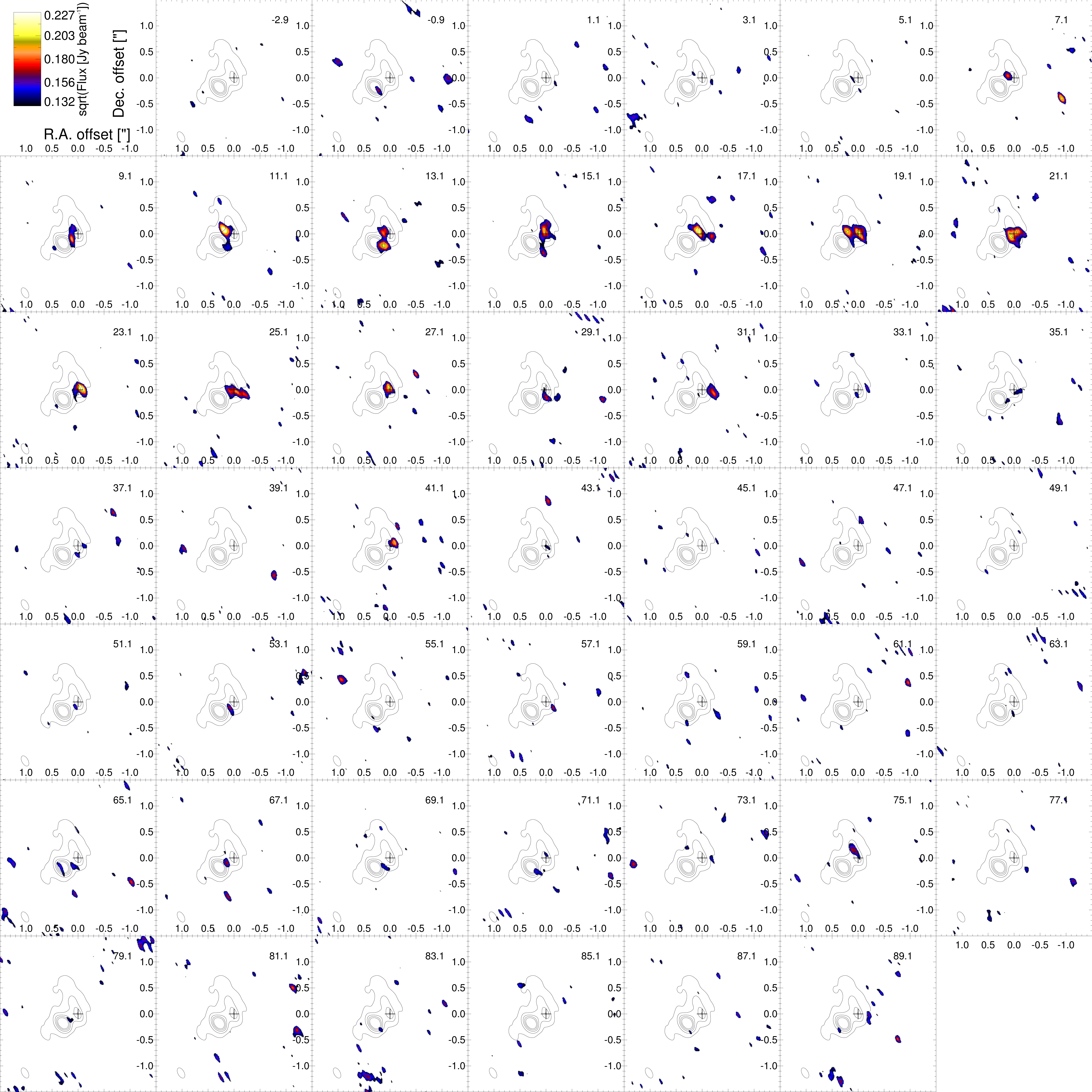}
\caption{Same as Fig.~\ref{fig:310554}, but for TiO$_2$ emission at 325.60\,GHz.}\label{fig:325601}
\end{figure*}

\begin{figure*}[p]
\centering
\subfigure[310.55\,GHz \label{fig:310554_mom0}]{\includegraphics[width=.245\linewidth]{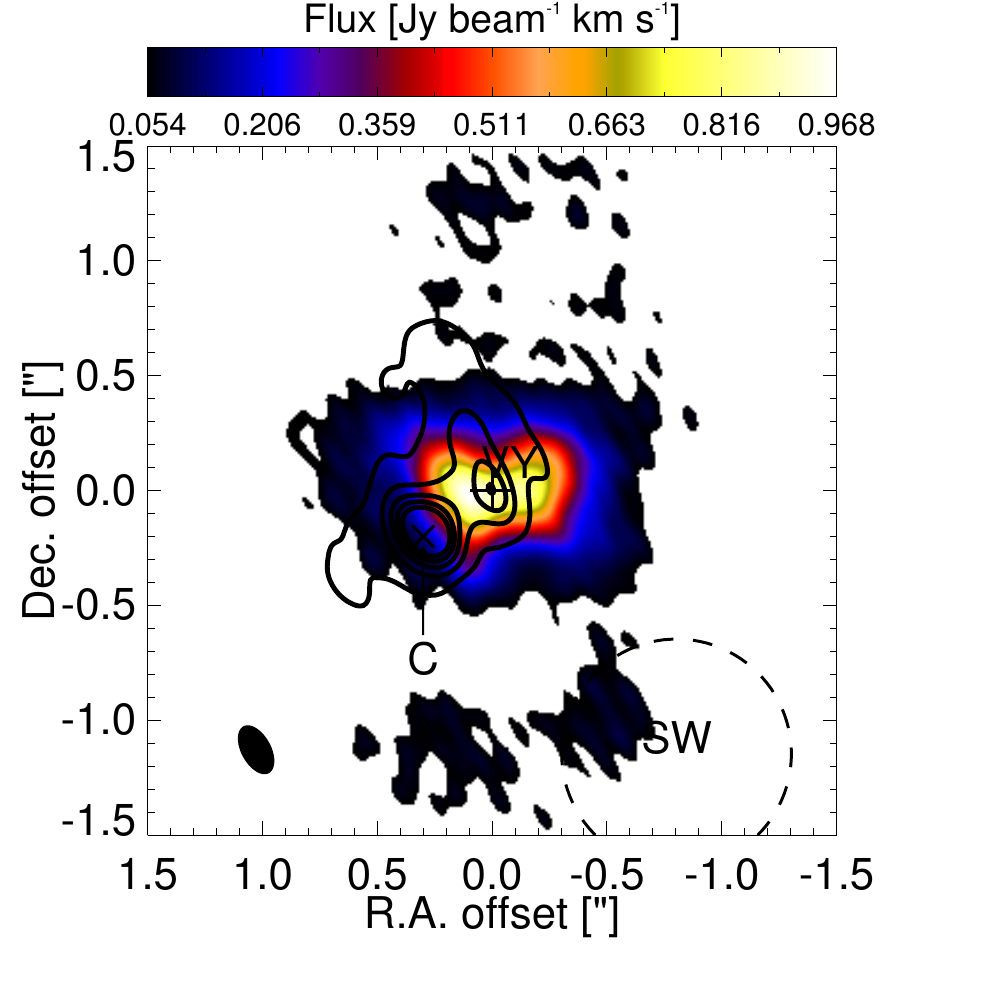}}
\subfigure[310.78\,GHz \label{fig:310783_mom0}]{\includegraphics[width=.245\linewidth]{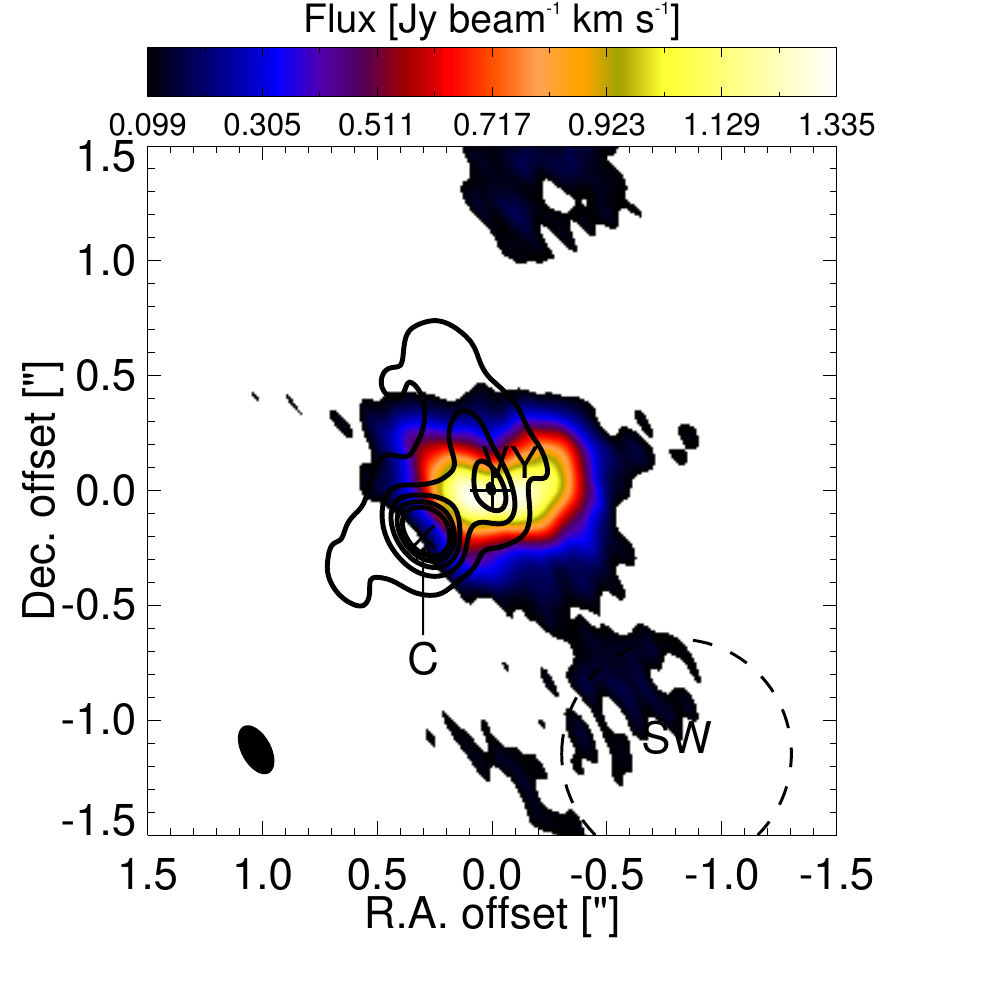}}
\subfigure[311.46\,GHz \label{fig:311462_mom0}]{\includegraphics[width=.245\linewidth]{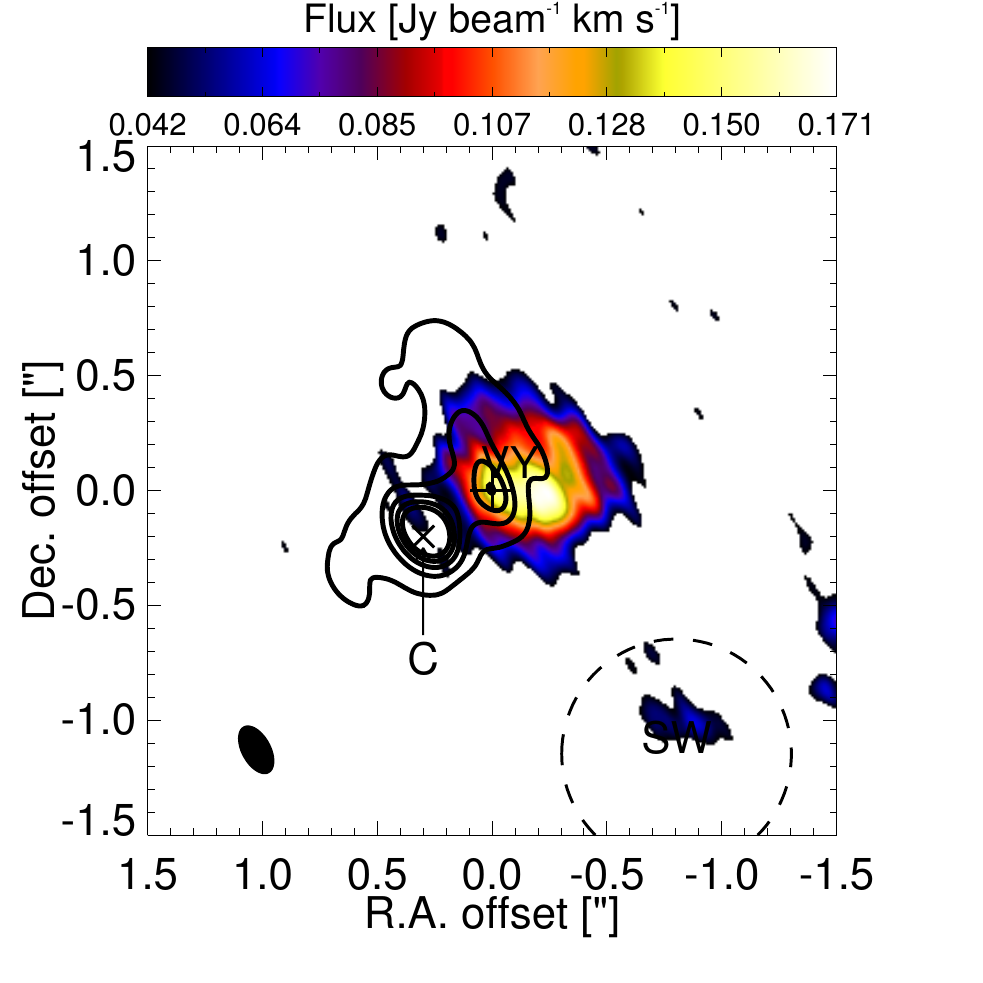}}
\subfigure[{312.25\,GHz}\label{fig:312248_mom0}]{\includegraphics[width=.245\linewidth]{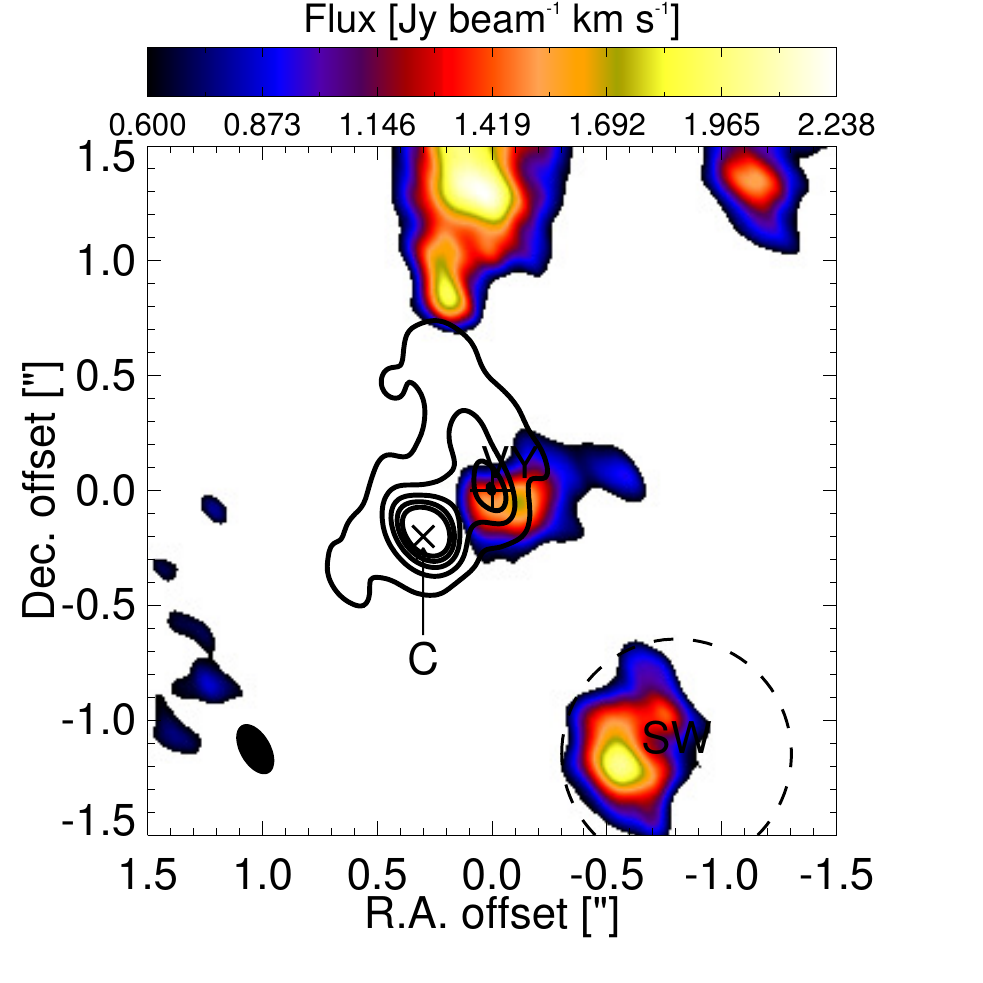}}
\subfigure[{312.73\,GHz}\label{fig:312732_mom0}]{\includegraphics[width=.245\linewidth]{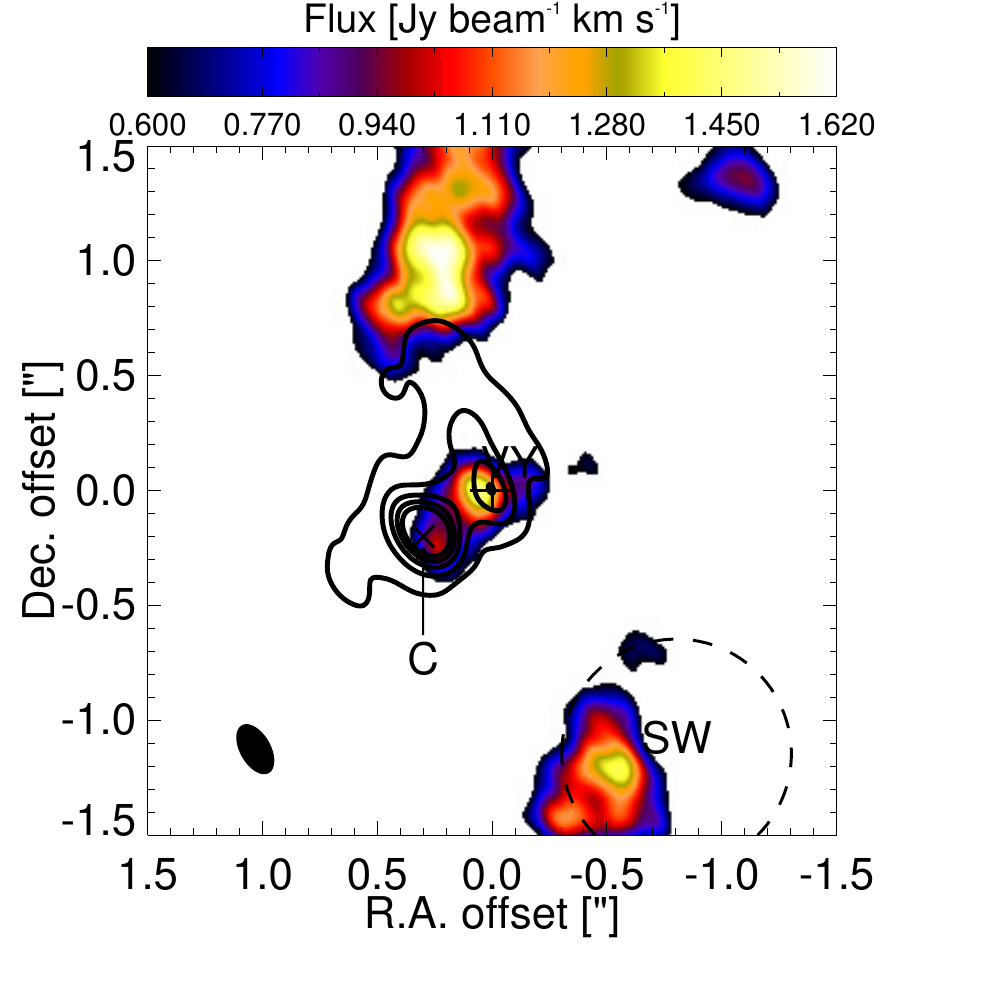}}
\subfigure[{312.82\,GHz}\label{fig:312817_mom0}]{\includegraphics[width=.245\linewidth]{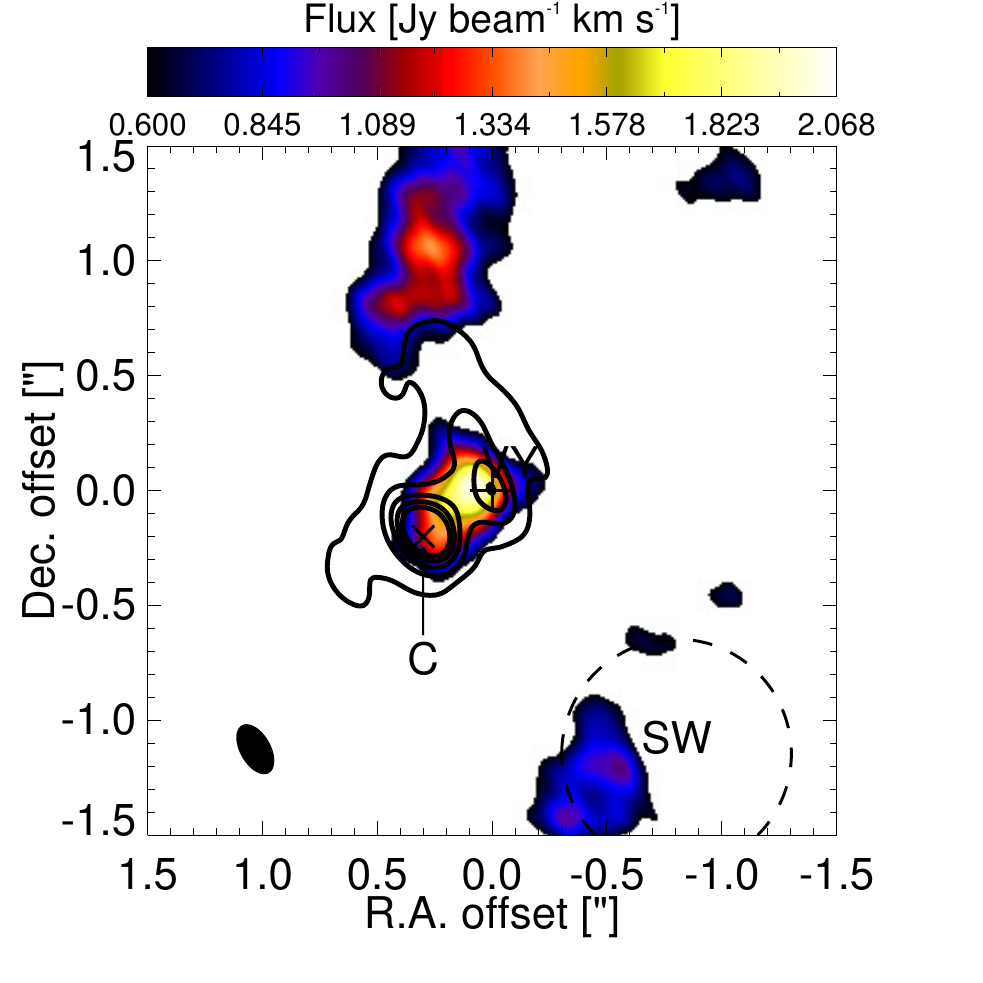}}
\subfigure[{321.40\,GHz.}\label{fig:321402_mom0}]{\includegraphics[width=.245\linewidth]{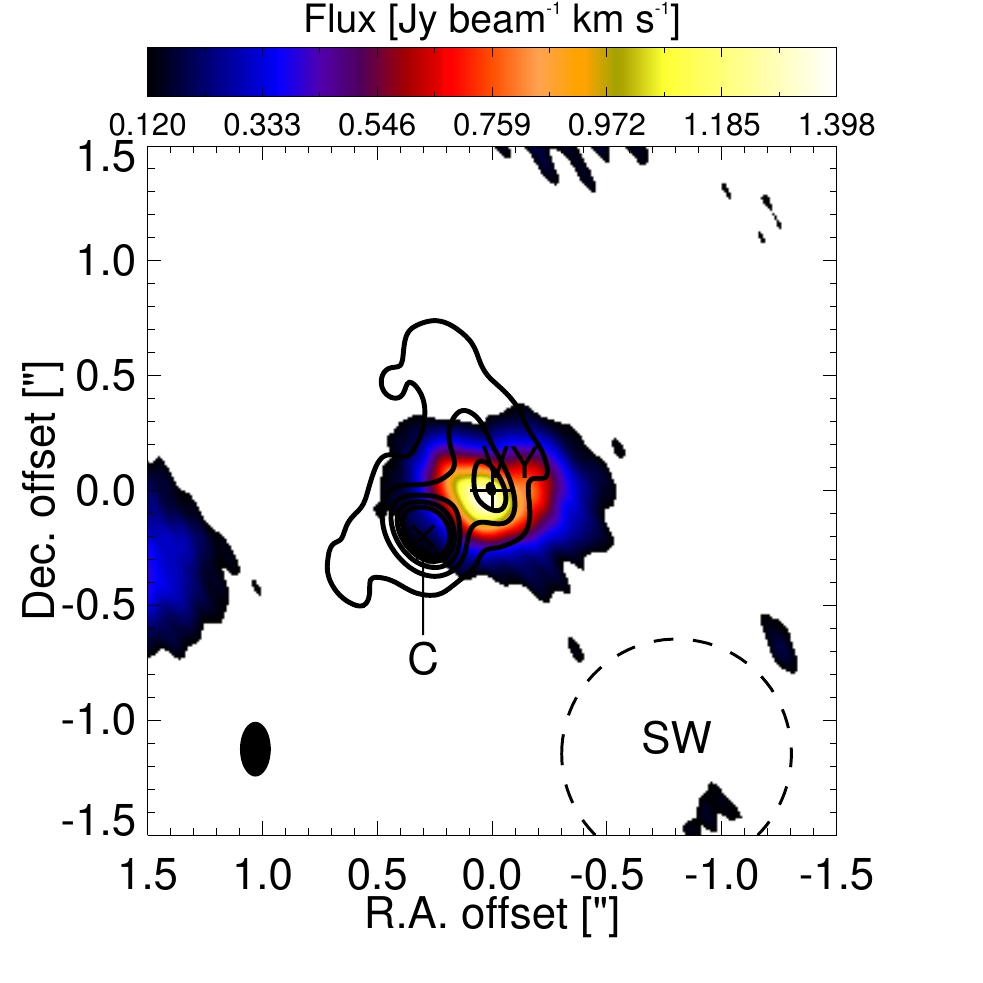}}
\subfigure[{321.50\,GHz}\label{fig:321501_mom0}]{\includegraphics[width=.245\linewidth]{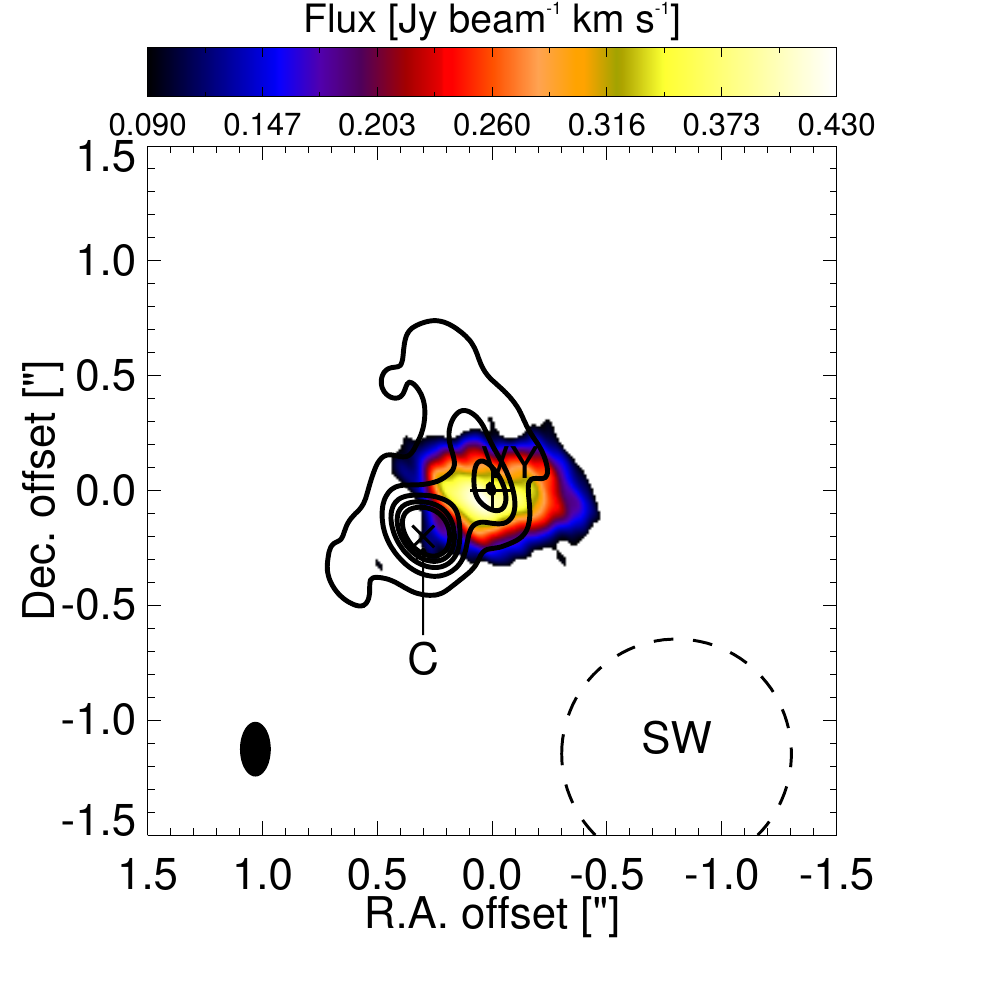}}
\subfigure[{322.33\,GHz}\label{fig:322334_mom0}]{\includegraphics[width=.245\linewidth]{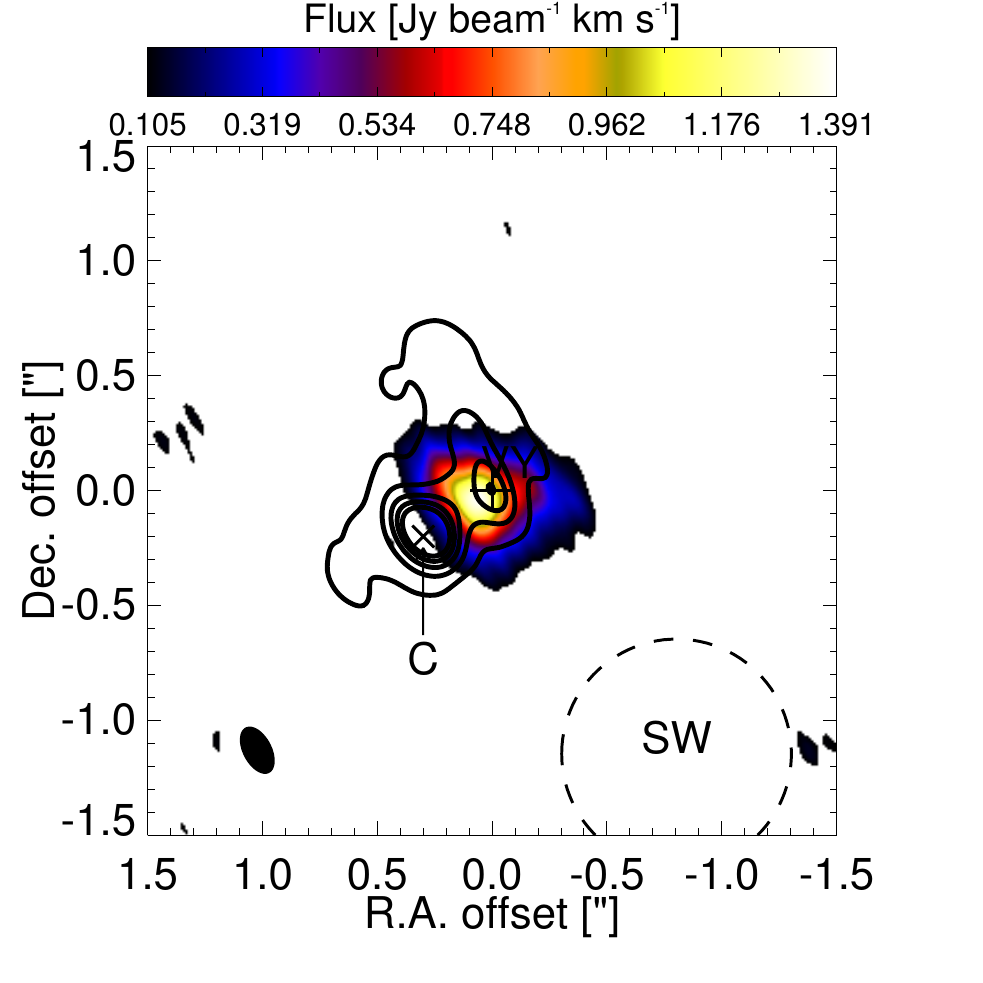}}
\subfigure[{322.61\,GHz}\label{fig:322613_mom0}]{\includegraphics[width=.245\linewidth]{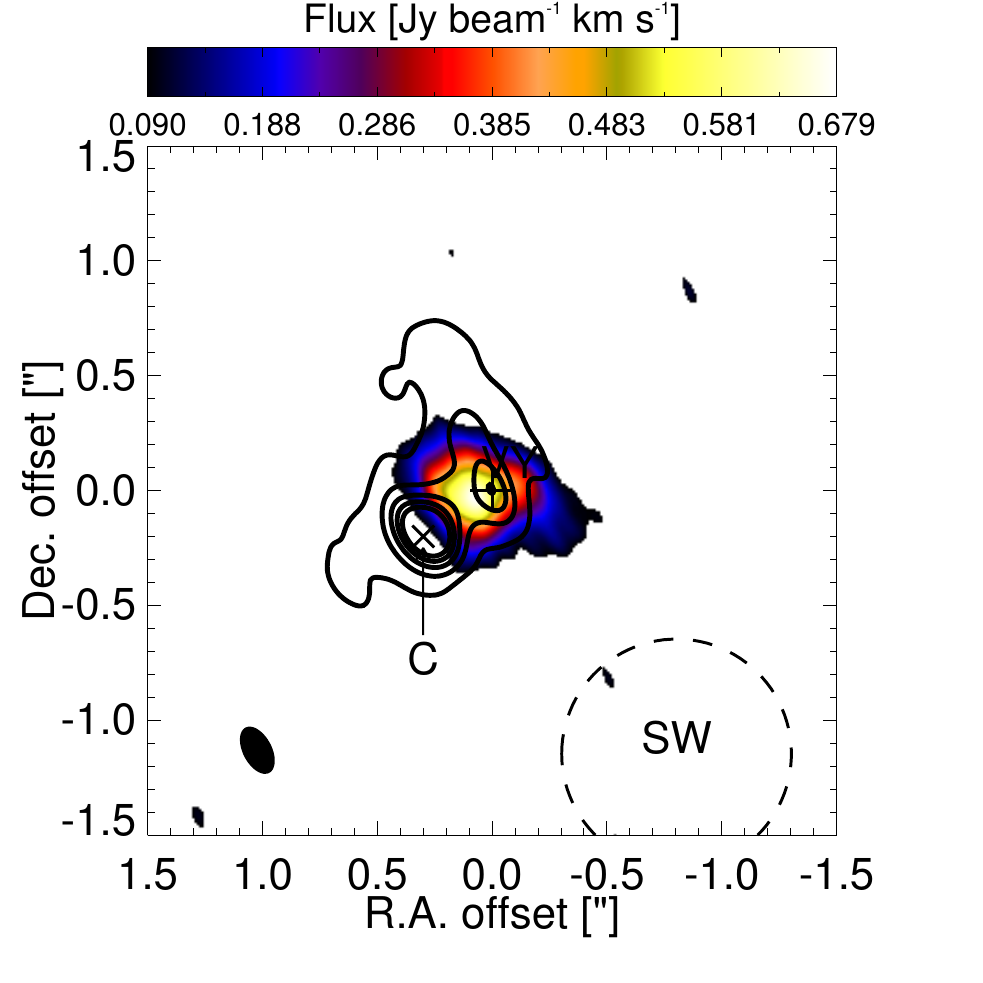}}
\subfigure[{324.49\,GHz}\label{fig:324493_mom0}]{\includegraphics[width=.245\linewidth]{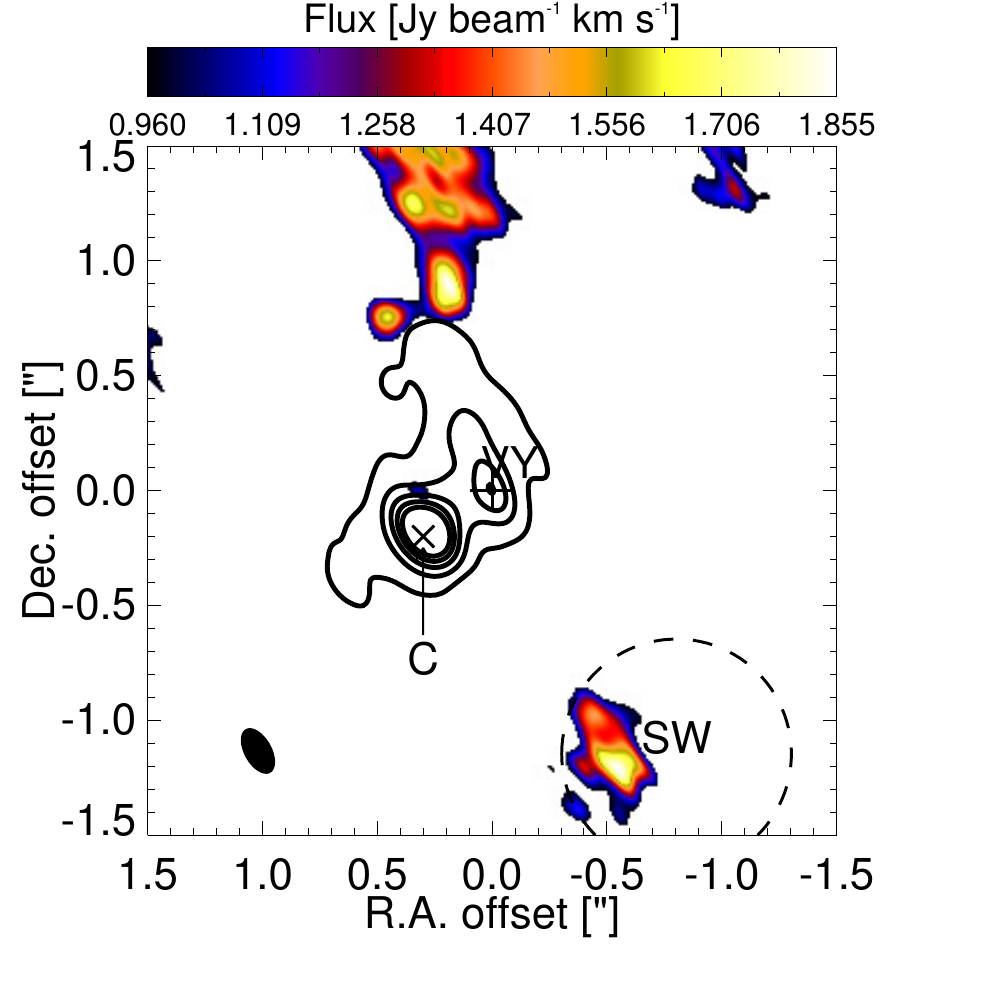}}
\subfigure[{324.97\,GHz}\label{fig:324966_mom0}]{\includegraphics[width=.245\linewidth]{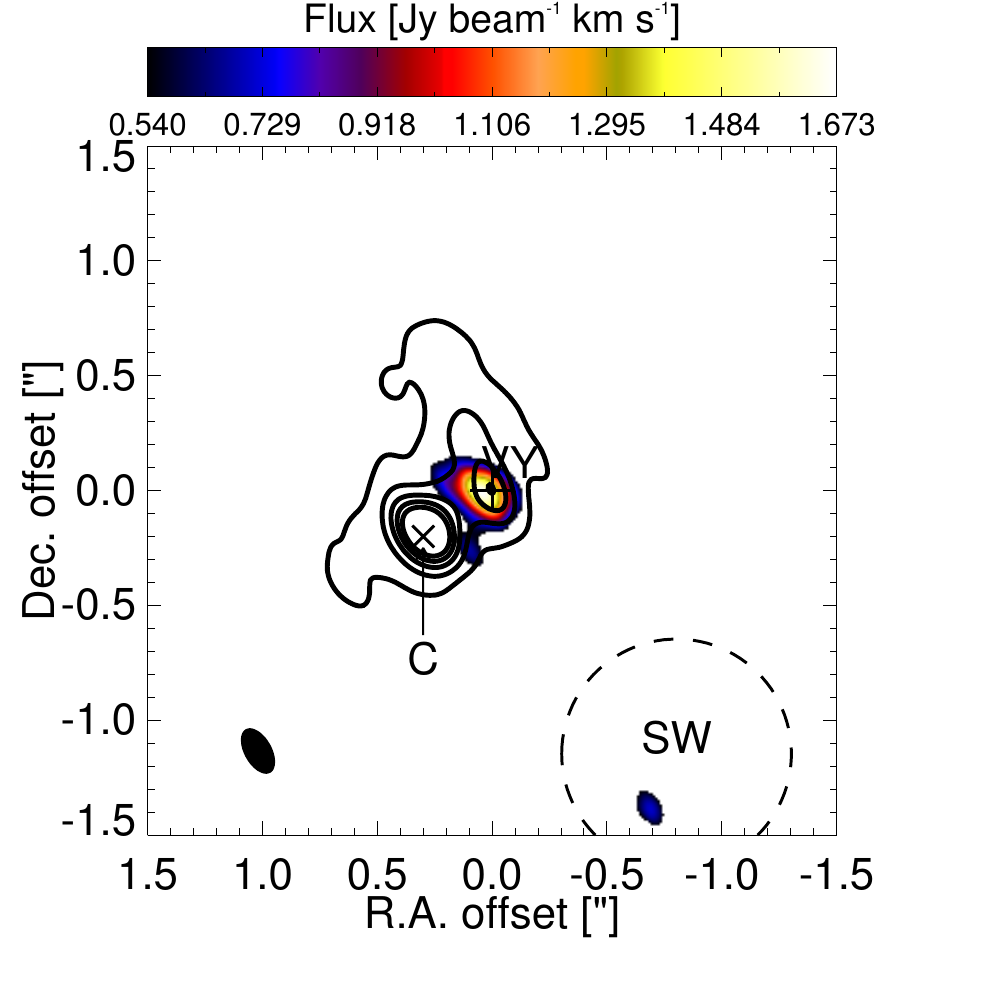}}
\subfigure[{325.32\,GHz}\label{fig:325322_mom0}]{\includegraphics[width=.245\linewidth]{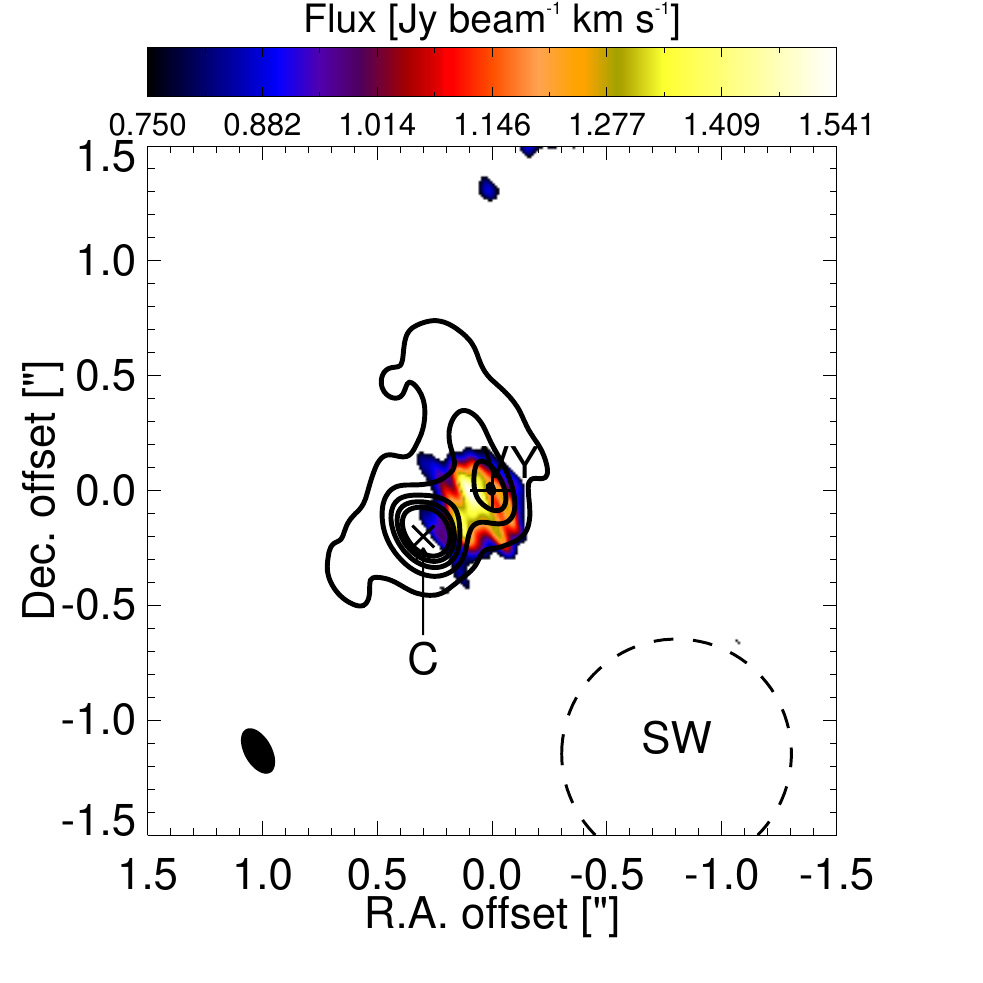}}
\subfigure[{325.50\,GHz}\label{fig:325500_mom0}]{\includegraphics[width=.245\linewidth]{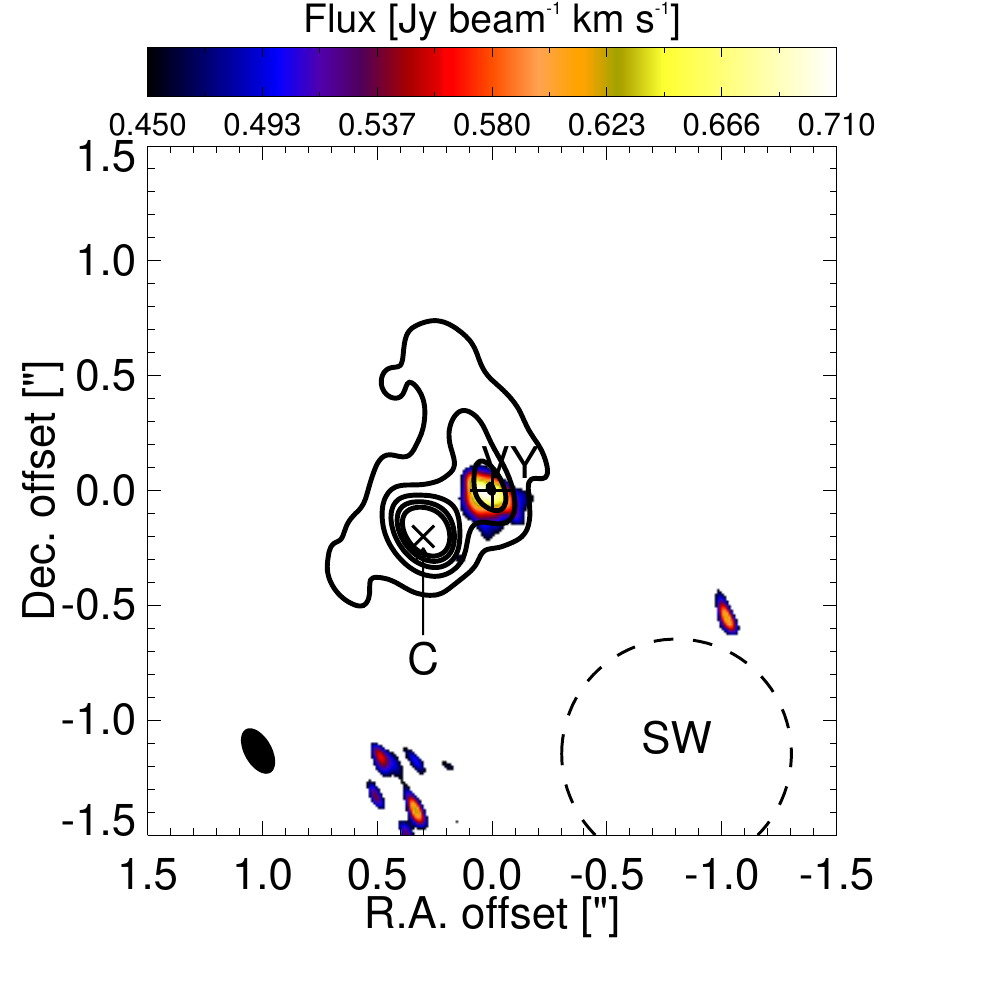}}
\subfigure[{325.60\,GHz}\label{fig:325601_mom0}]{\includegraphics[width=.245\linewidth]{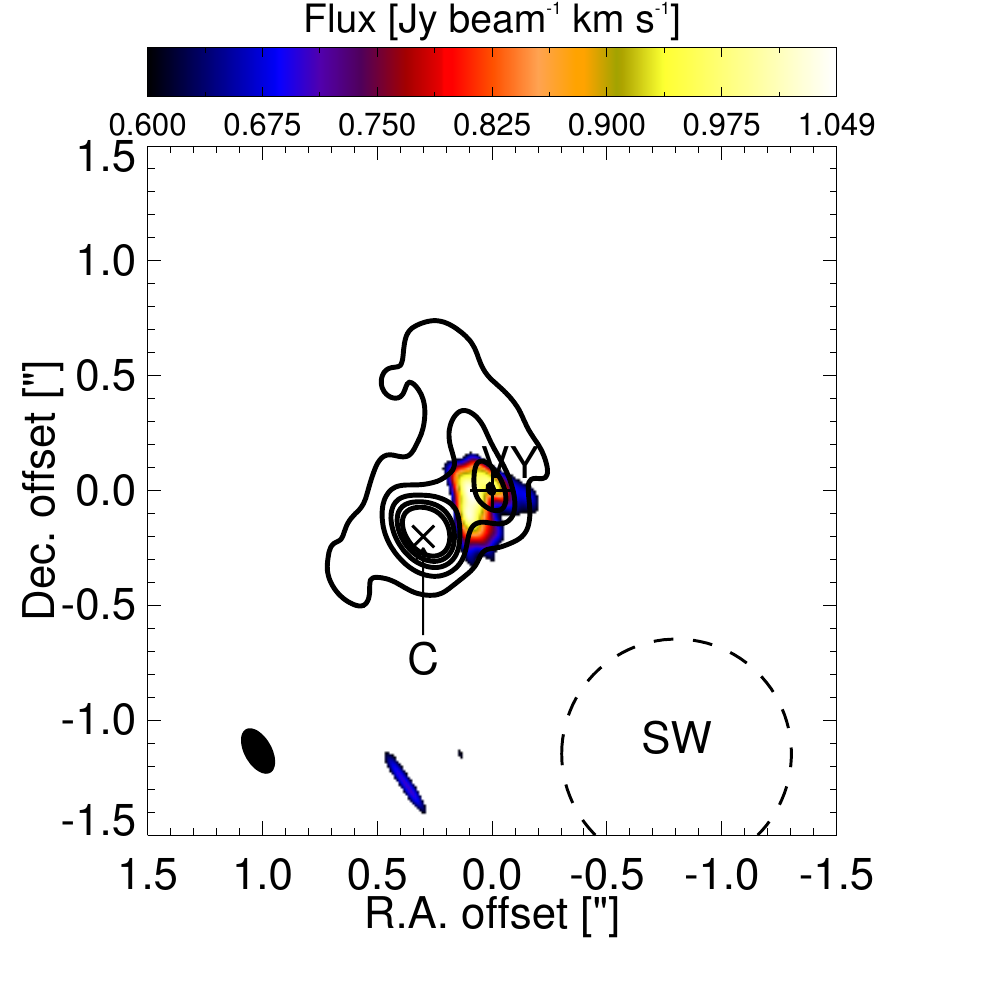}}
\caption{Integrated-intensity maps of \tioo emission. Colour maps show the intensity integrated over the \vlsr-ranges indicated in Table~\ref{tbl:lineID} and Fig.~\ref{fig:tio2all} and cut off at $3\sigma$. Contours show the ALMA 321\,GHz continuum. Labels indicate the positions of the star (+, VY) and the continuum component (x, C) to the south-east \citepalias{ogorman2015_alma_vycma,richards2014_alma_vycma}, and the position and approximate extent of the south-west clump of \citet[][SW and a dashed 1\arcsec\ diameter circle]{shenoy2013_vycma_AO_2to5micron}. The presence of cleaning artefacts, mainly at $\sim$312\,GHz, is addressed in Appendix~\ref{app:channels}.\label{fig:mom0_all}}
\end{figure*}

\begin{figure*}[p]
\subfigure[TiO$_2$ at 310.55\,GHz \label{fig:310554_SW}]{\includegraphics[width=.245\linewidth]{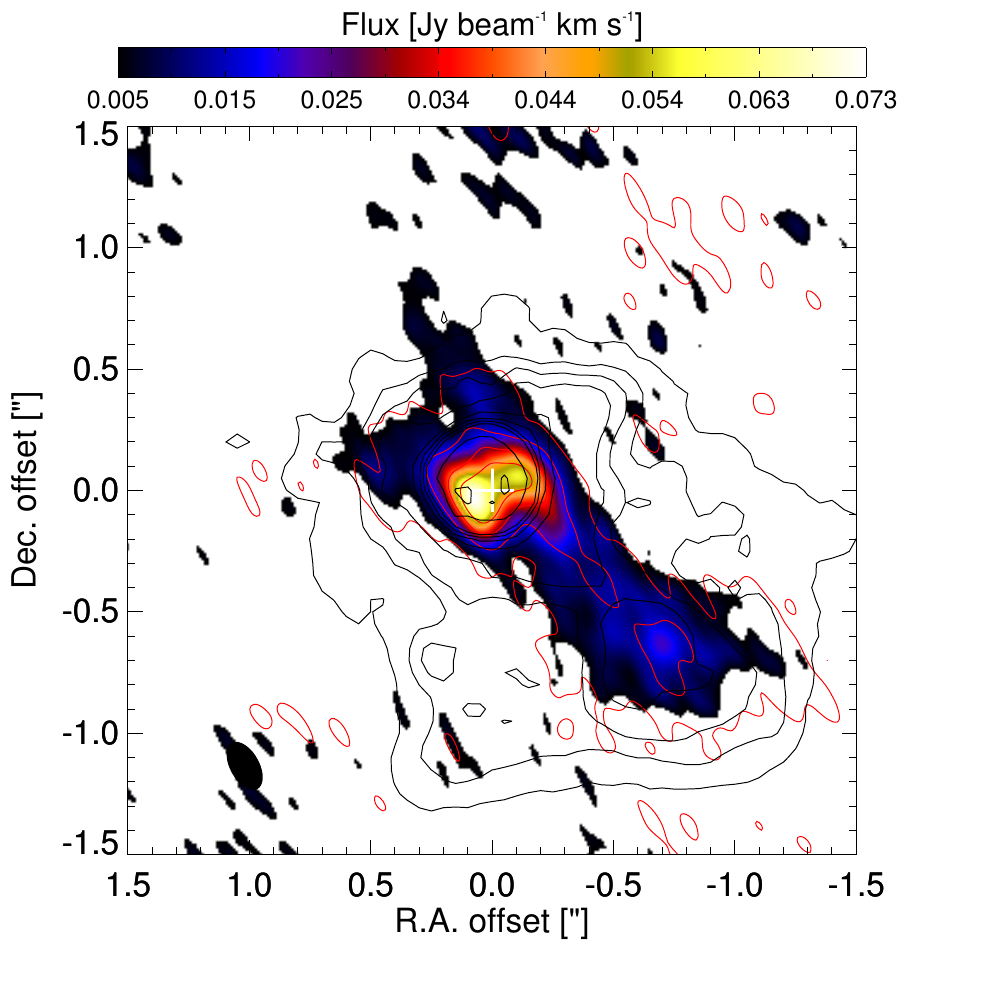}}
\subfigure[{TiO$_2$ at 312.25\,GHz}\label{fig:312248_SW}]{\includegraphics[width=.245\linewidth]{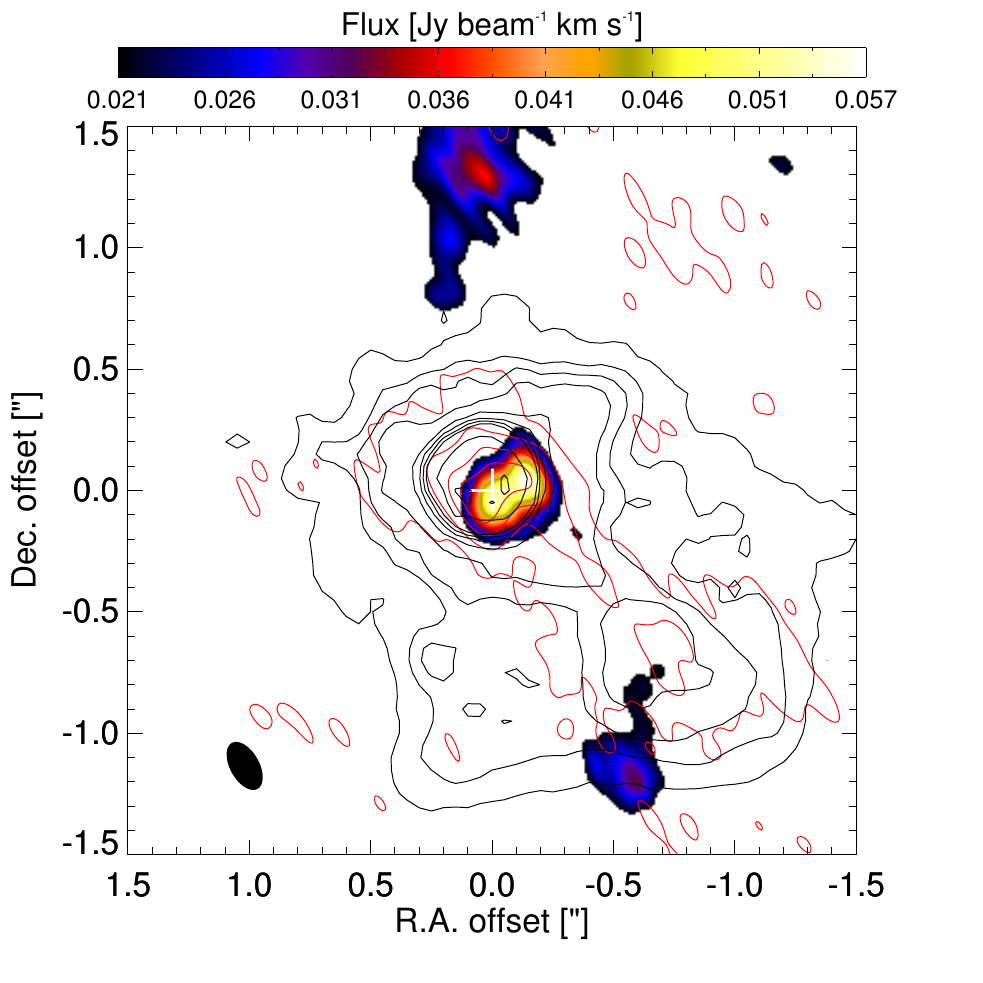}}
\subfigure[{TiO$_2$ at 312.73\,GHz}\label{fig:312732_SW}]{\includegraphics[width=.245\linewidth]{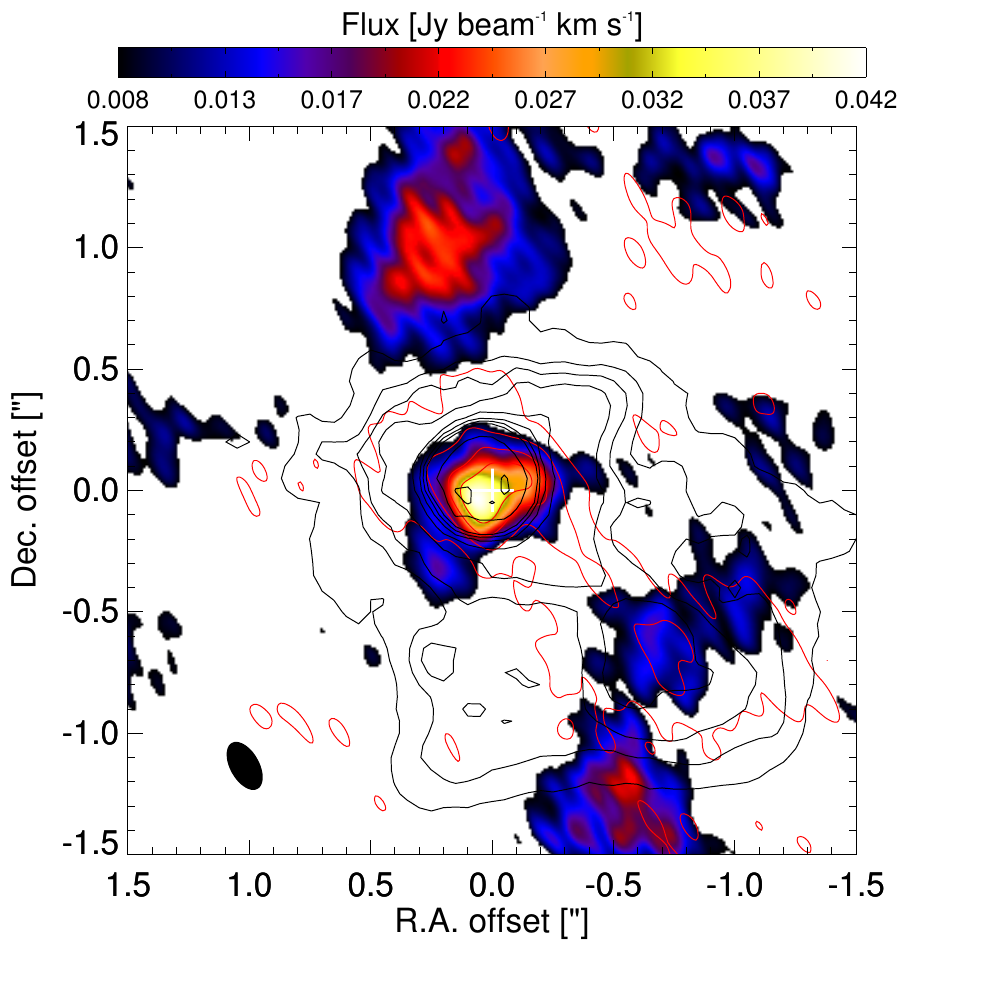}}
\subfigure[{TiO$_2$ at 312.82\,GHz}\label{fig:312817_SW}]{\includegraphics[width=.245\linewidth]{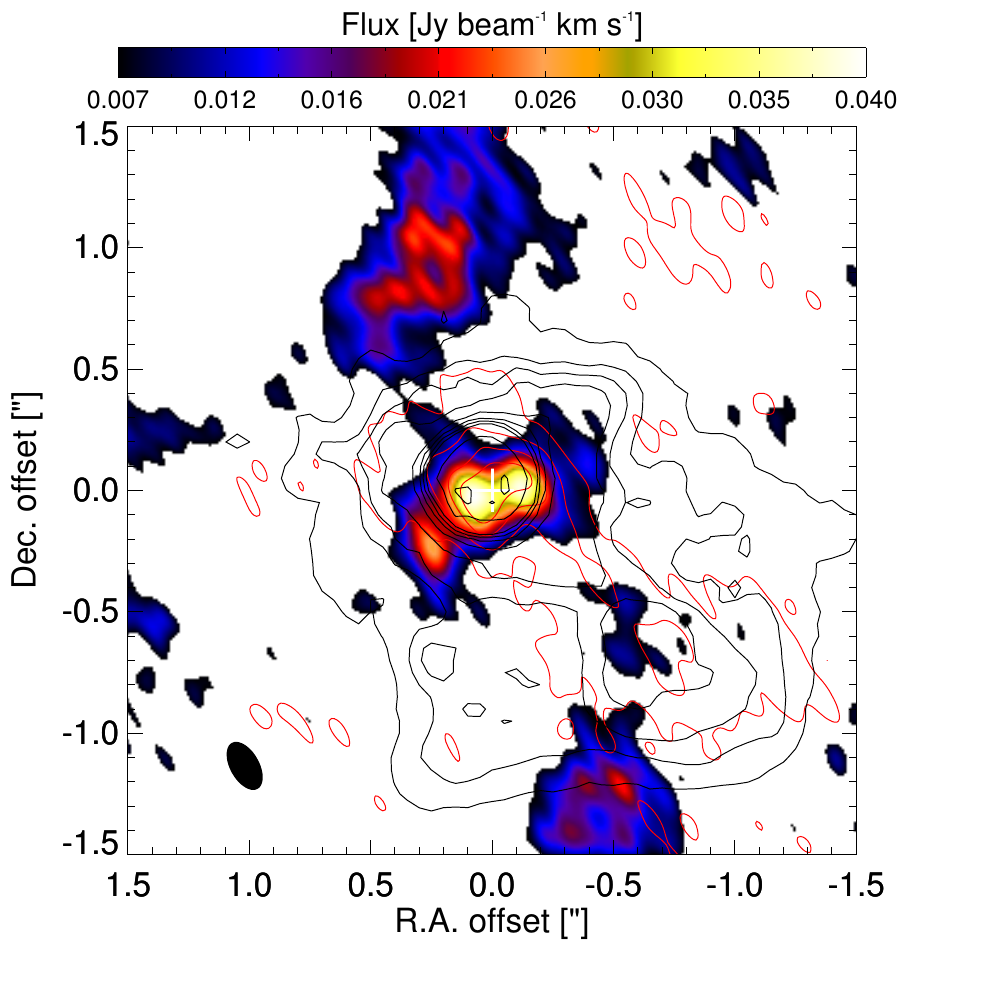}}
\subfigure[{TiO$_2$ at 321.40\,GHz.}\label{fig:321402_SW}]{\includegraphics[width=.245\linewidth]{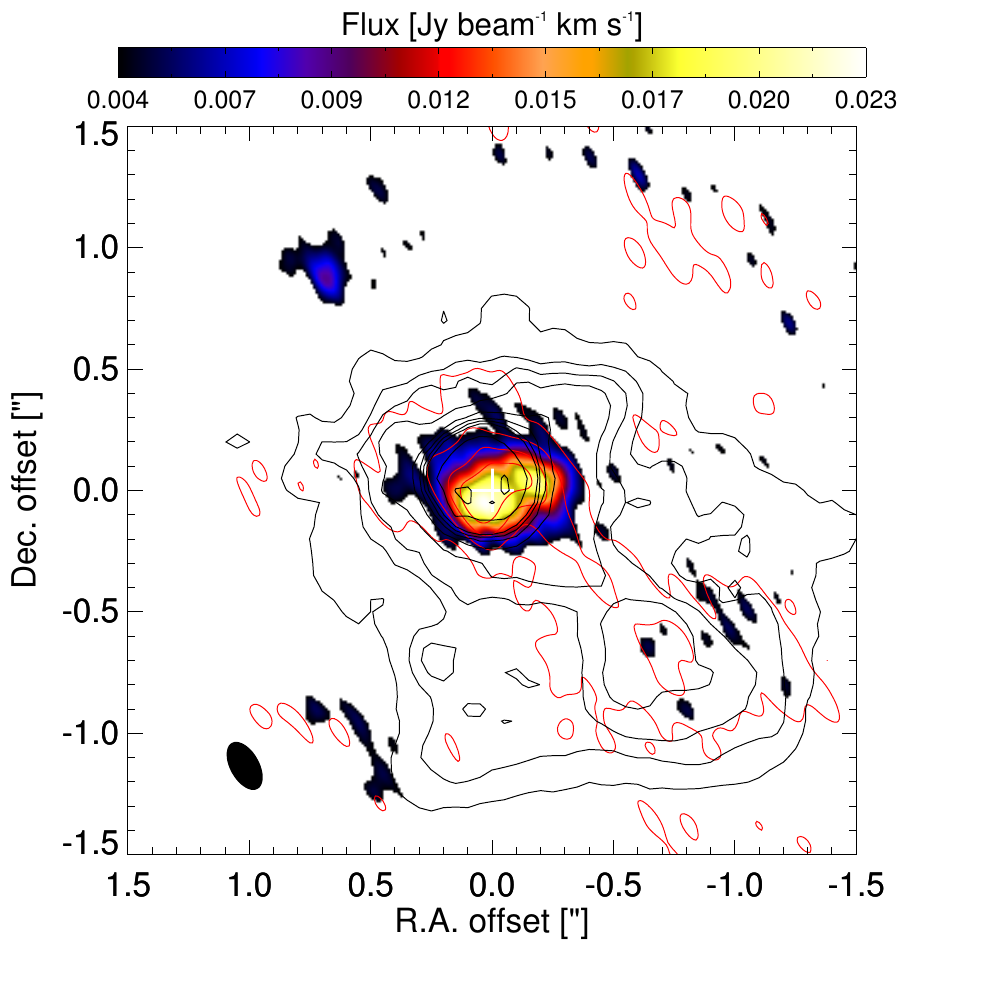}}
\subfigure[{TiO$_2$ at 321.50\,GHz}\label{fig:321501_SW}]{\includegraphics[width=.245\linewidth]{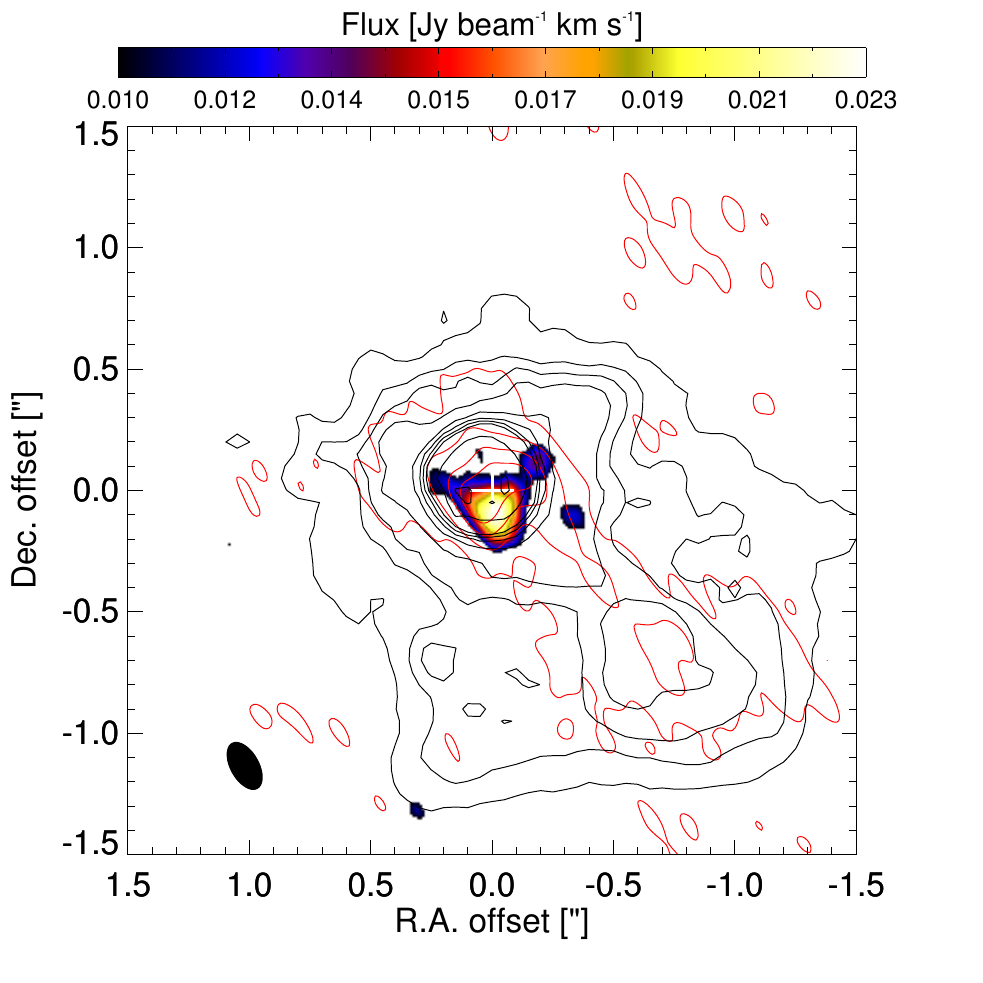}}
\subfigure[{TiO$_2$ at 322.33\,GHz}\label{fig:322334_SW}]{\includegraphics[width=.245\linewidth]{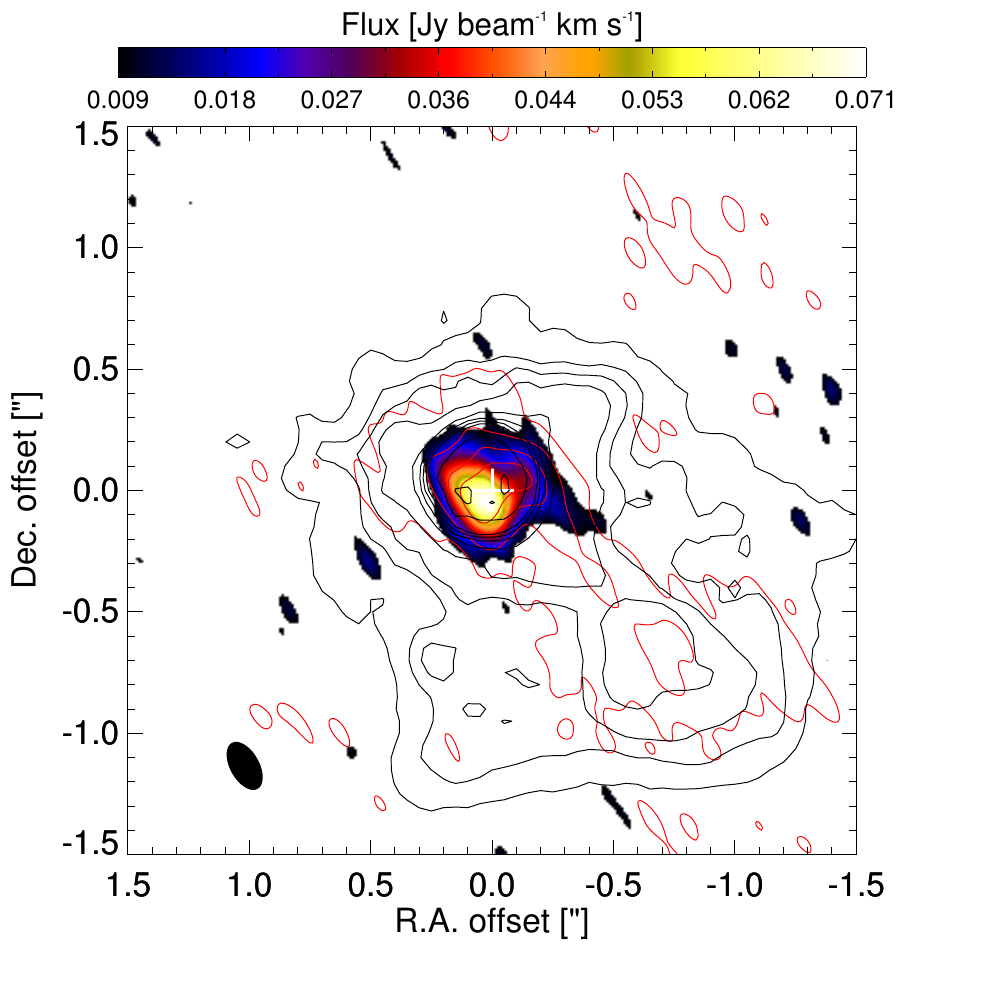}}
\subfigure[{TiO$_2$ at 322.61\,GHz}\label{fig:322613_SW}]{\includegraphics[width=.245\linewidth]{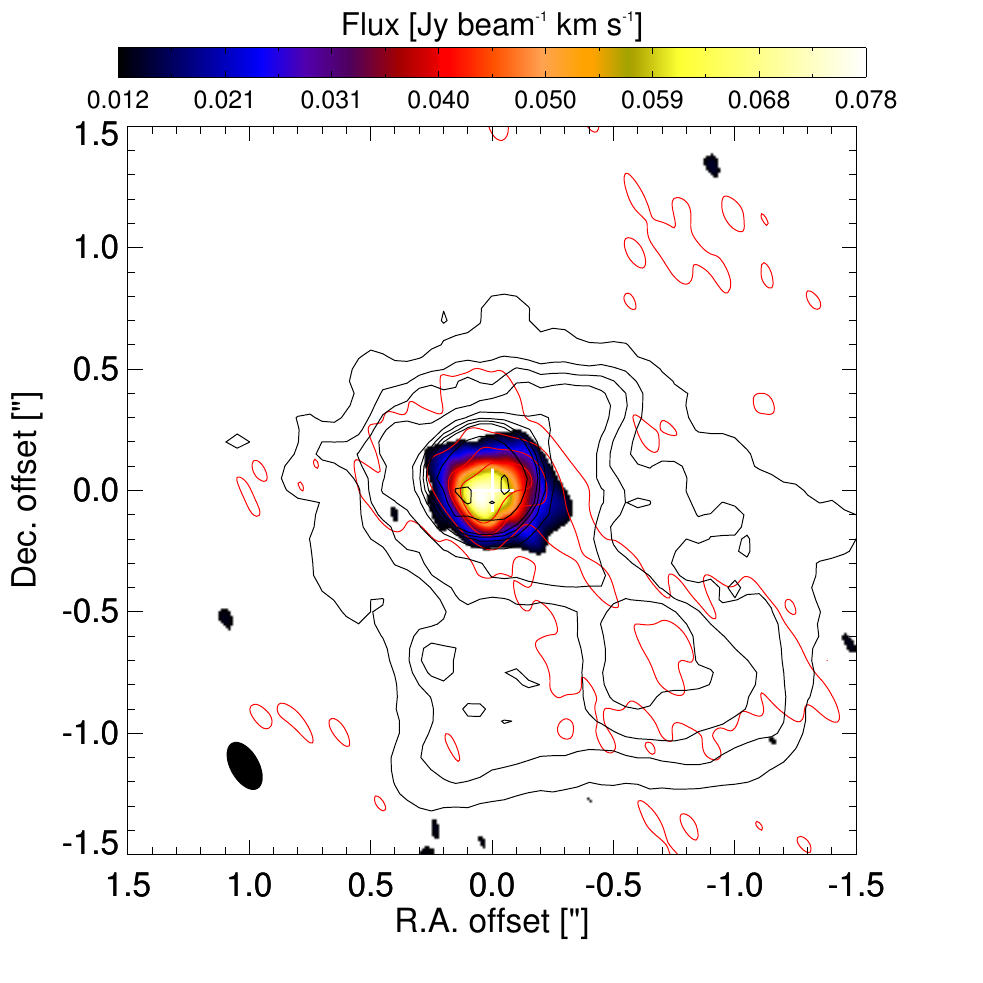}}
\subfigure[{TiO$_2$ at 324.49\,GHz}\label{fig:324493_SW}]{\includegraphics[width=.245\linewidth]{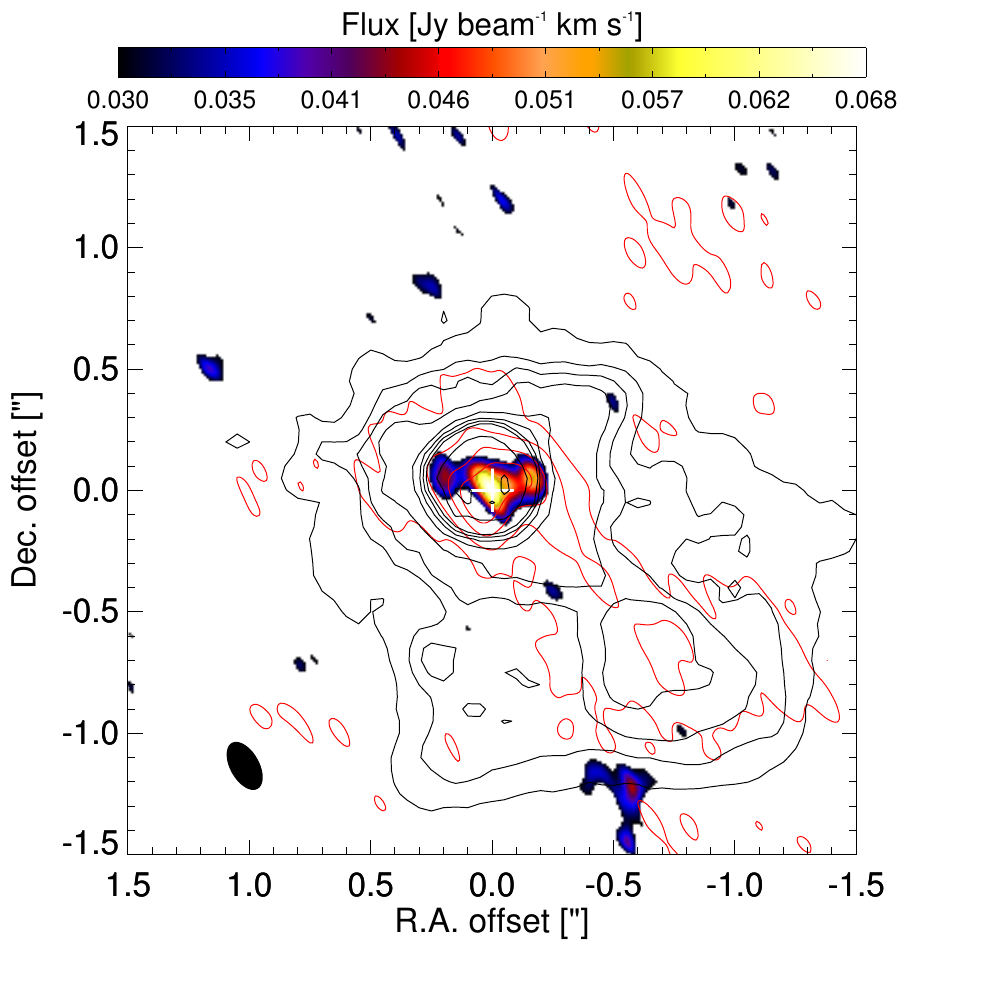}}
\subfigure[{TiO$_2$ at 324.97\,GHz}\label{fig:324966_SW}]{\includegraphics[width=.245\linewidth]{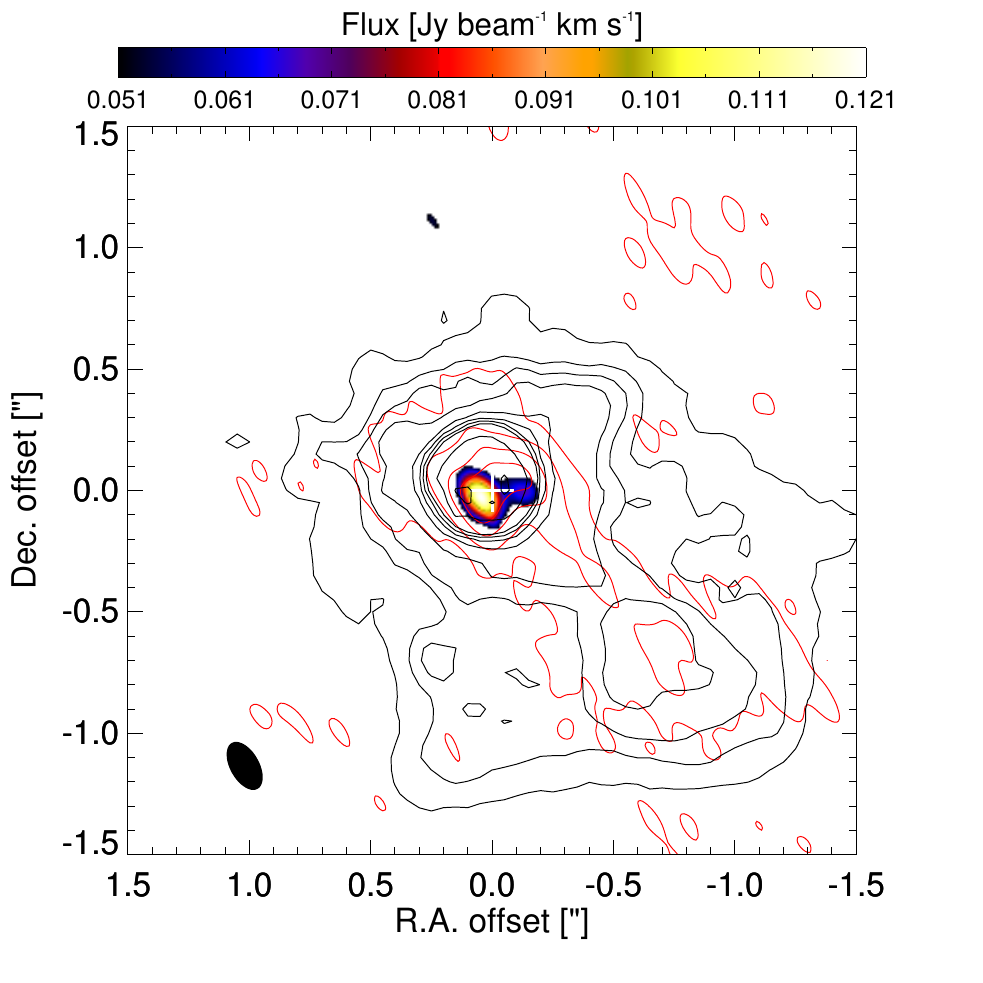}}
\subfigure[{TiO$_2$ at 325.32\,GHz}\label{fig:325322_SW}]{\includegraphics[width=.245\linewidth]{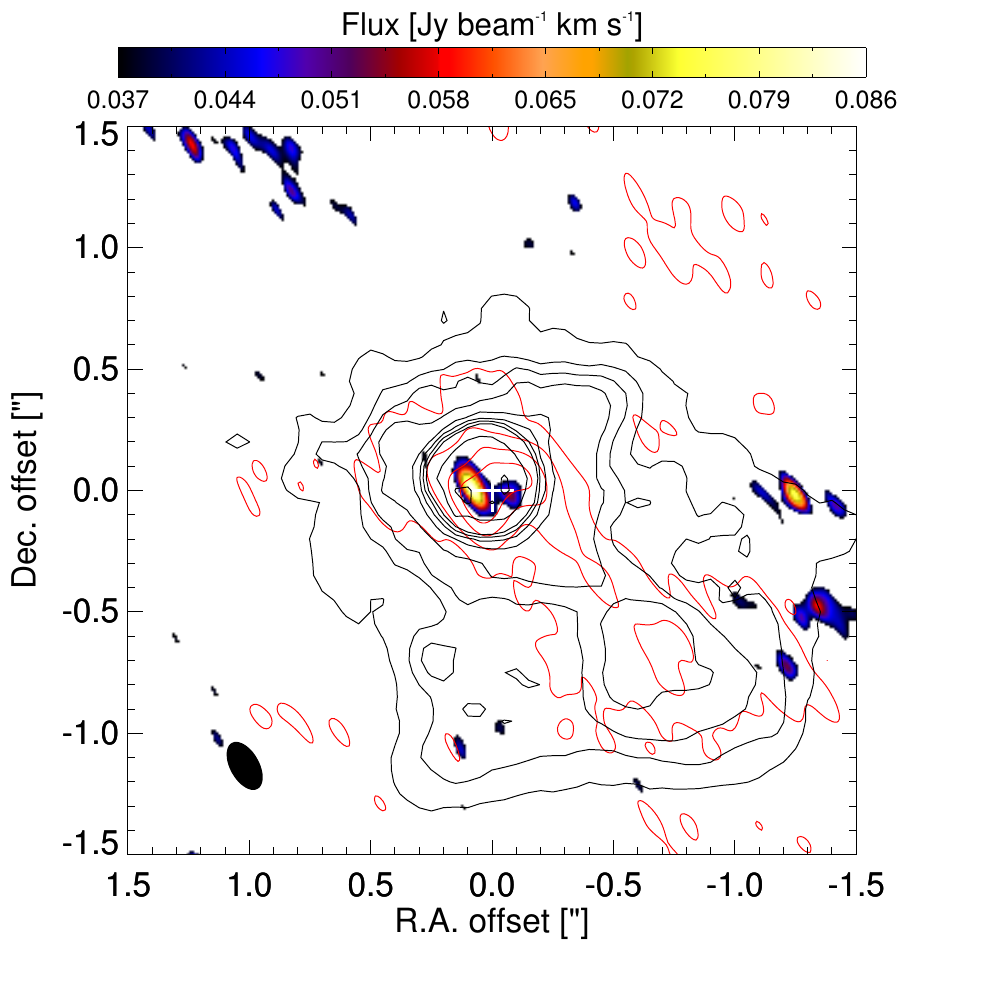}}
\subfigure[{TiO$_2$ at 325.50\,GHz}\label{fig:325500_SW}]{\includegraphics[width=.245\linewidth]{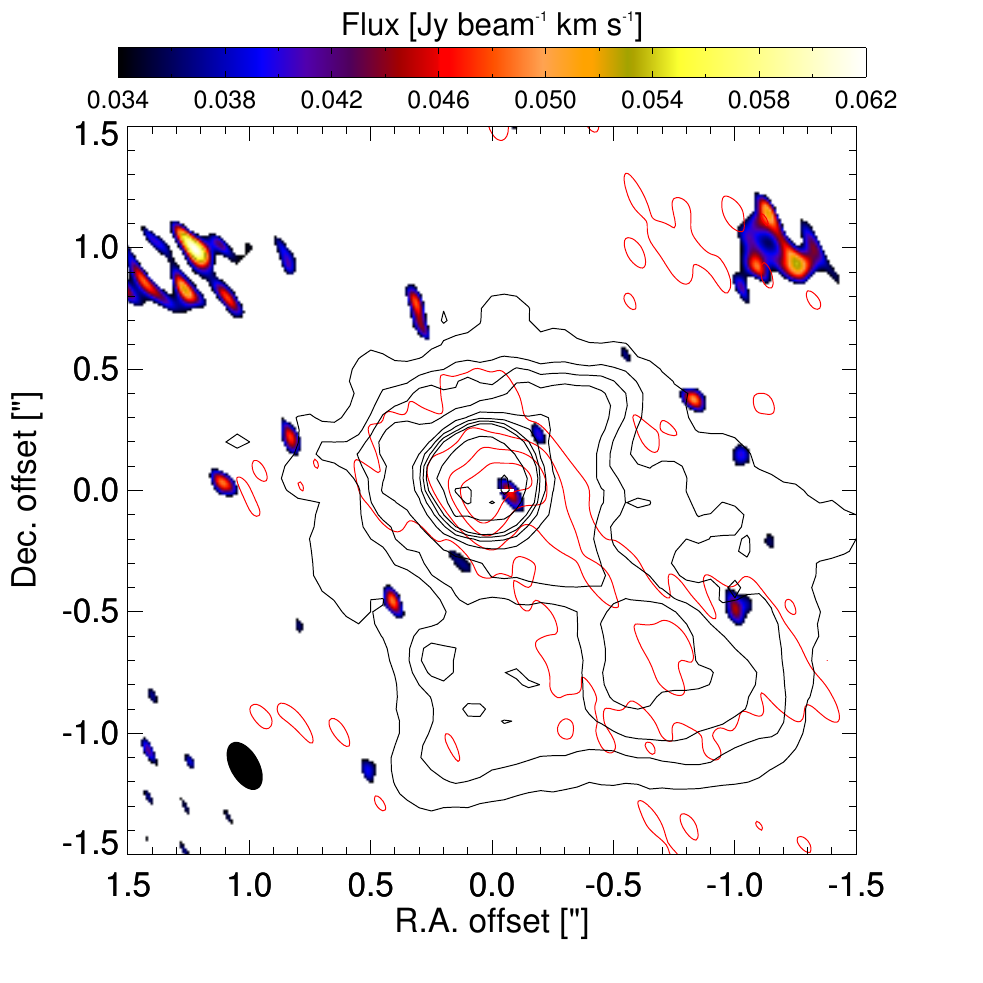}}
\subfigure[{TiO$_2$ at 325.60\,GHz}\label{fig:325601_SW}]{\includegraphics[width=.245\linewidth]{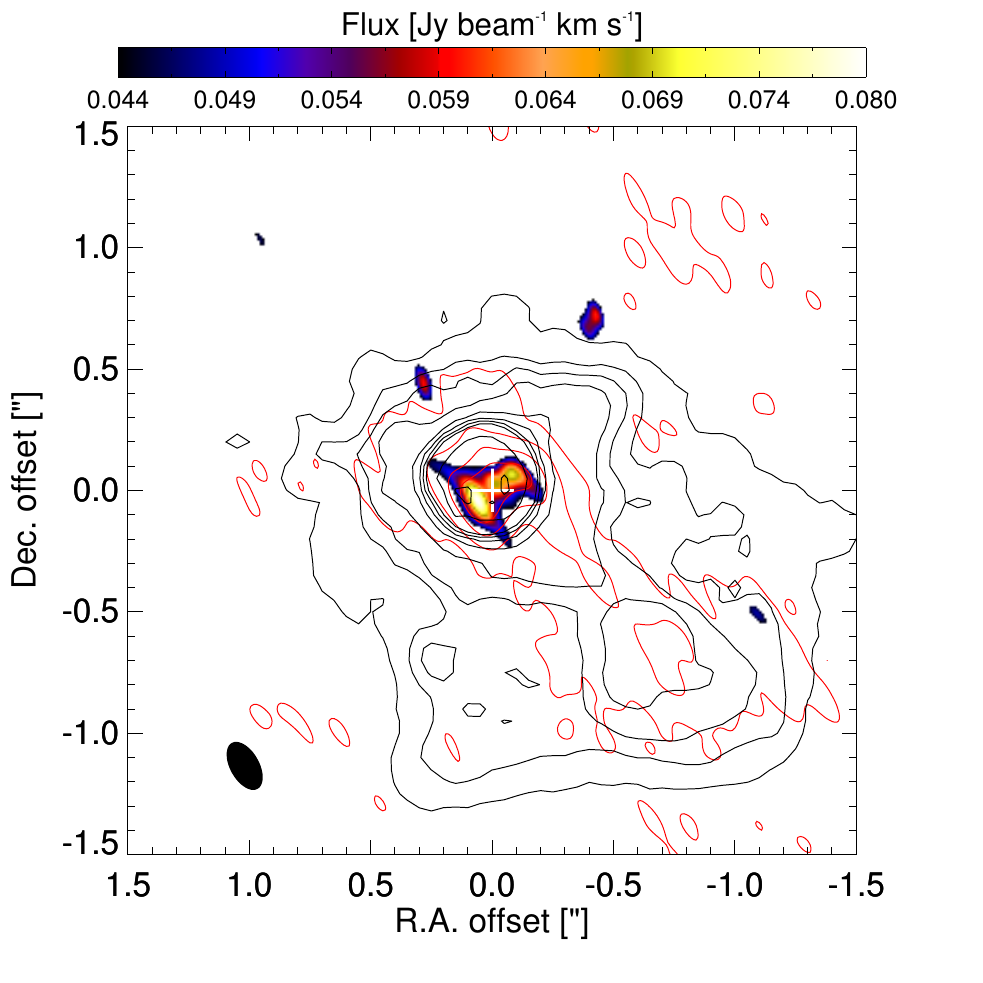}}
\caption{Comparison of TiO$_2$ emission lines listed in Table~\ref{tbl:lineID} to south-west tail at 310.78\,GHz (red contours) and HST image (black contours). Emission integrated over $19\leq\vlsr\leq22$\,\kms plotted at $>3\sigma$ (colour scale). Emission of 311.46\,GHz was omitted since the relevant \vlsr-range is not covered by the observations.\label{fig:SW_all}}\end{figure*}

Figures~\ref{fig:310554} to \ref{fig:325601} show channel maps of the detected TiO$_2$ emission lines (Table~\ref{tbl:lineID}, Fig.~\ref{fig:tio2all}) at a velocity resolution of 2\,\kms, covering the range $-14\leq\vlsr\leq78$\,\kms. This \vlsr-range covers the bulk of all \tioo emission; emission at more extreme velocities is no longer visible in the channel maps. Figure~\ref{fig:mom0_all} shows integrated-intensity maps for all listed lines, covering the \vlsr-ranges indicated in Table~\ref{tbl:lineID}. Figure~\ref{fig:SW_all} shows a comparison of the \tioo line emission to the south-west tail detected at 310.78\,GHz, which is discussed in Sects.~\ref{sect:morphology} and \ref{sect:swtail}.

\clearpage
\clearpage
\clearpage
\clearpage
\clearpage
\clearpage
\clearpage
\clearpage
\clearpage
\clearpage
\clearpage
\clearpage
\clearpage
\clearpage

\end{appendix}
\end{document}